\documentclass[twocolumn]{aastex62}

\newcommand\msun{M_{\odot}}
\newcommand\mhi{M_{HI}}
\newcommand\LV{L_{V}}
\newcommand\lsun{L_{\odot}}
\newcommand\vsys{V_{sys}}
\newcommand\wfty{W_{50}}
\newcommand{\kms}{{\ensuremath{\mathrm{km\,s^{-1}}}}}
\newcommand\mhilim{M^{lim}_{HI}}

\shorttitle{LSB Dwarfs Candidates in HI}
\shortauthors{Karunakaran et al.}

\begin{document}

\title{Neutral Hydrogen Observations of Low Surface Brightness Galaxies around M101 and NGC 5485}

\correspondingauthor{Ananthan Karunakaran}
\email{a.karunakaran@queensu.ca}

\author[0000-0001-8855-3635]{A. Karunakaran}
\affiliation{Department of Physics, Engineering Physics and Astronomy, Queen’s University, Kingston, ON K7L 3N6, Canada}
\author[0000-0002-0956-7949]{K. Spekkens}
\affiliation{Department of Physics and Space Science, Royal Military College of Canada P.O. Box 17000, Station Forces Kingston, ON K7K 7B4, Canada}
\affiliation{Department of Physics, Engineering Physics and Astronomy, Queen’s University, Kingston, ON K7L 3N6, Canada}

\author[0000-0001-8354-7279]{P. Bennet}
\affiliation{Physics \& Astronomy Department, Texas Tech University, Box 41051, Lubbock, TX 79409-1051, USA}
\author[0000-0003-4102-380X]{D. J. Sand}
\affiliation{Steward Observatory, University of Arizona, 933 North Cherry Avenue, Rm. N204, Tucson, AZ 85721-0065, USA}

\author[0000-0002-1763-4128]{D. Crnojevi\'c}
\affiliation{University of Tampa, 401 West Kennedy Boulevard, Tampa, FL 33606, USA}
\affiliation{Physics \& Astronomy Department, Texas Tech University, Box 41051, Lubbock, TX 79409-1051, USA}
\author[0000-0002-5177-727X]{D. Zaritsky}
\affiliation{Steward Observatory, University of Arizona, 933 North Cherry Avenue, Rm. N204, Tucson, AZ 85721-0065, USA}

\begin{abstract}We present atomic hydrogen (HI) observations using the Robert C. Byrd Green Bank Telescope along the lines-of-sight to 27 low surface brightness (LSB) dwarf galaxy candidates discovered in optical searches around M101. We detect HI reservoirs in 5 targets and place stringent upper limits on the remaining 22, implying that they are gas poor. The distances to our HI detections range from 7 Mpc --150 Mpc, demonstrating the utility of wide-bandpass HI observations as a follow-up tool. The systemic velocities of 3 detections are consistent with that of the NGC~5485 group behind M101, and we suggest that our 15 non-detections with lower distance limits from the optical are associated with and have been stripped by that group. We find that the gas richnesses of confirmed M101 satellites are broadly consistent with those of the Milky Way satellites, as well as with those of satellites around other hosts of comparable mass, when survey completeness is taken into account. This suggests that satellite quenching and gas stripping proceeds similarly around halos of similar mass, in line with theoretical expectations.\end{abstract}


\section{Introduction} \label{sec:intro}
Recent improvements in astronomical instrumentation \citep[e.g.][]{DragonflyArrayOriginal,HSC-SurveyOverview} and novel image searching algorithms \citep[e.g.][]{Paulspaper,LeoILeoTriplet,SMUDGes} have revitalized studies of the low surface brightness (LSB) universe.\ Many of these LSB features detected resemble dwarf galaxies \citep{2014MerritLSBsM101}, while some have extreme properties relative to the high surface brightness galaxy population \citep[e.g.][]{UDGsvandokkum}.

Wide-field LSB searches for faint companions of nearby galaxies allow comparisons with Local Group satellite populations, and there have been concerted efforts to obtain a census of the satellite populations around such hosts \citep{M81Groupmembers,M81Confirmed,TBG-Kara2014,2014MerritLSBsM101,2015AstBu..70..379K,PISCeS-CenAsats,JavanmardiDGSAT,Paulspaper,Mullerspaper,M94sats,Crnojevi__2019}. One focus is to measure the satellite populations of Milky Way-like hosts \citep{JavanmardiDGSAT,GehaSAGA} in order to constrain cosmic variance among halos of similar mass \citep{Cosmicvariance,2019fieldercosmicvariance}. In this context, the Satellites Around Galactic Analogues survey \citep[SAGA,][]{GehaSAGA} has revealed a population of star-forming satellites which project within the virial radii of Milky Way-like hosts. The on-going star formation in the bulk of the SAGA detections stands in contrast to the largely quiescent, gas-poor satellite population of the Milky Way itself \citep[e.g.][]{lgcatalog,HIMilkyWay}. This raises the possibility that satellite gas stripping and star formation quenching proceed differently around halos of similar mass, providing an important new observational constraint on the underlying physics \citep[e.g.][]{Wetzel2015,Environmentalquenching2018,RPS-SimpsonAuriga,2019garrisonkimmel}.

The nearby Milky Way-like spiral M101 \citep[$D_{M101} = 7.0$ Mpc, $V_{M101} = 241 \, \kms$ adopted in this work;][]{LeeandJang2012,Tikhonov,m101vsys} has been the target of several LSB searches aiming to characterize its satellite populations \citep{TBG-Kara2014,2014MerritLSBsM101,2015AstBu..70..379K,JavanmardiDGSAT,Paulspaper,Mullerspaper}.\ Many of the detected LSB objects have morphologies consistent with satellites of that host, although their asymmetric sky distribution is challenging to explain if they are all bona-fide companions \citep{Paulspaper}.

This tension is alleviated somewhat by the presence of a background group in the vicinity of M101 \citep[$D_{BG}\sim 27 \,\mathrm{Mpc}, \, V_{BG}\sim 1961 \, \kms$;][]{2016ApJ...833..168M,Galaxygroups2massTully}, which includes the massive ellipticals NGC~5485 and NGC~5473 among its $\sim25$ members \citep{karaandmakaM101,SaulderNGC5485}. Indeed, follow-up observations to characterize the properties of the LSB dwarf candidates around M101 suggest that many lie at the distance of the NGC~5485 group \citep{2016ApJ...833..168M,Paulm101hst,CarlstenSBFM101}. The distances to many others remain unconstrained, and there is a need for additional follow-up observations to constrain both their locations and their physical properties.

Deep searches for the atomic hydrogen (HI) reservoirs of LSB detections are a powerful tool for constraining their physical properties \citep{HCGUDGs}: detections provide estimates of distance and HI mass, while non-detections yield upper limits on the gas richnesses of objects along the line-of-sight (LOS). Since the HI content of field dwarfs \citep{huangalfalafa,Bradford2015MAIN} as well as the environmental dependence of that content \citep{HILocalGroup,HIMilkyWay,2017browngasstripping} are well-characterized in the local universe, HI follow-up observations provide a mechanism for constraining the physical properties of the LSB dwarf candidates and the impact of their environments on those properties.

We present HI follow-up observations along the LOS to 27 LSB dwarf candidates in the M101 region using the Robert C. Byrd Green Bank Telescope (GBT).\ We aim to characterize their gas properties and to use that information to place the M101 and NGC~5485 systems into context with other Milky Way-like hosts and nearby galaxy groups.

The structure of this paper is as follows.\ In Section \ref{sec:sample}, we describe our HI target selection. We outline our observations and data reduction procedure in Section \ref{sec:Obsanddata}. In Section \ref{sec:results}, we present the properties of our detections as well as upper limits on the gas content of our non-detections.\ In Section \ref{sec:Discussion}, we discuss the implications of these findings for the membership of the M101 and NGC~5485 group systems, and for the influence of these hosts on the gas content of the satellites. We summarize in Section \ref{sec:Conclusion}.

\section{Sample Selection} \label{sec:sample}

The M101 satellite system consists of 4 previously known, more luminous satellites \citep[$M_V < -14$: NGC~5474, NGC~5477, Holm IV, and DDO 194;][]{Tikhonov}. In their HI mapping study of the M101 region, \citet{Mihosm101HI} detected two low-mass HI features. We do not consider these in our sample selection: one does not have an optical counterpart, while the other has a recessional velocity that is not consistent with M101. Recent optical LSB searches, described in the previous section, have identified fainter satellite candidates \citep[5 of which have been recently confirmed with $M_V < -8.2$: DF-1, DF-2, DF-3, DwA, Dw9;][see below]{2017M101LuminosityFn,Paulm101hst}. As described below, we select our sample of HI targets from M101 satellite candidates identified in those LSB searches.

\citet[][hereafter M14]{2014MerritLSBsM101} carried out one of the first modern LSB searches around M101 using the Dragonfly Telephoto Array, discovering 7 dwarf candidates in a $\sim 9 \, \mathrm{deg^2}$ field.\ In their survey around nearby spirals, \citet[][hereafter K15]{2015AstBu..70..379K} discovered an additional 4 dwarf candidates in the region around M101, one of which (DwA/DGSAT1) was separately discovered by \citet{JavanmardiDGSAT}. LSB searches were also conducted using extant data. \citet[][hereafter B17]{Paulspaper} confirmed the M14 and K15 detections and discovered 38 new LSB dwarf candidates in a $\sim 9 \, \mathrm{deg^2}$ field around M101 in CFHT Legacy Survey data using a semi-automated algorithm tuned to reveal LSB features. While the preceding searches focused in the immediate region around M101, \citet[][hereafter M17]{Mullerspaper} searched through SDSS data for new LSB dwarf candidates over $330 \, \mathrm{deg^2}$ in the broader M101 group complex to reveal 6 new candidates beyond the M101 virial radius. In combination, these 4 studies revealed a total of 55 unique LSB dwarf candidates in the region around M101.

We select all LSB dwarf candidates from the combined samples of M14, K15, B17, and M17 with apparent magnitudes $m_V < 19.5\, \mathrm{mag}$ for HI follow-up with the GBT. The optical properties of the resulting 27 targets are given in Table \ref{table:maintable}. Since we determine observing times from a gas richness scaling relation for local dwarfs \citep[][see Section \ref{sec:Obsanddata}]{Bradford2015MAIN}, our adopted $m_V$ threshold implies a follow-up time for each target of a few hours at most (see column 10 of Table \ref{table:maintable}). Column 7 of Table \ref{table:maintable} gives the references for the optical properties that we adopt in this study.

A  variety of studies have constrained distances to the M101 LSB dwarf candidates since their discovery, and columns 8 and 9 of Table \ref{table:maintable} list the value that we adopt for our follow-up targets.\ HST campaigns reported by \citet[][hereafter M16]{2016ApJ...833..168M}, \citet[][hereafter D17]{2017M101LuminosityFn} and \citet[][hereafter B19]{Paulm101hst} have either confirmed a dwarf candidate's association with M101 from Tip of the Red Giant Branch (TRGB) distances from resolved star colour-magnitude diagrams, or reported lower distance limits derived from the lack of resolved stars.\ Distance constraints are also reported by \citet[][hereafter C19]{CarlstenSBFM101} via a surface brightness fluctuation (SBF) technique. We adopt distances or lower limits from M16, D17 or B19 when available, and otherwise we adopt the C19 estimates.\ Table \ref{table:maintable} shows that of the 27 M101 dwarf candidates that we targeted  4 have distance estimates consistent with M101, 18 have distance lower limits that place them in the M101 background, and the remaining 5 have no prior distance information. 

We note that although a combination of stellar mass and colour is a more accurate predictor of gas content than stellar mass alone for the high surface brightness galaxy population \citep[e.g.][]{GASS,Brown2015}, we do not use $g-r$ as a selection criterion for our HI follow-up sample.\ Instead, we investigate the utility of the available colours for LSB dwarf candidates as a predictor of gas richness in Section \ref{sec:Discussion}.

\section{Observations and Data Reduction} \label{sec:Obsanddata}


We performed 63 hours of observations between 2016 August and 2019 August using the GBT to determine the HI content along the LOS to the 27 LSB dwarf candidates in Table \ref{table:maintable} (programs AGBT16B-046, AGBT17A-188, and AGBT17B-235). We used the L-band receiver and the Versatile GBT Astronomical Spectrometer (VEGAS) with a bandpass of 100 MHz and spectral resolution of 3.1 kHz, dumping data every 5 seconds to build up the requisite integration time.\ This wide bandpass allows for the HI spectral line to be detected at heliocentric velocities up to $\sim 14000 \, \kms$. This observational setup was used for all but one candidate, DF-3, which had a distance estimate that is consistent with that of M101 (D17) at the time it was observed.\ For this target, we used a narrower bandpass, 11.72 MHz, and a higher spectral resolution, 0.4 kHz, centered at the heliocentric velocity of M101, $V_{M101} = 241 \, \kms$.  

Our observations spent an equivalent amount of time on the target of interest (i.e. the ``ON") and on a set of reference positions (that collectively constitute the ``OFF"). We estimate the integration times for our targets using $m_V$ calculated from $m_g$ and $g-r$ following the relations in \citet{Jestervband} to reach a gas richness of $\frac{\mhi}{\LV} \sim$ 1 $\frac{\msun}{\lsun}$ ($\sim0.5$ dex below the \citealt{Bradford2015MAIN} scaling relations at $\LV \lesssim 10^{9} \lsun$) with $S/N = 5$ in a single $25 \, \kms$ channel. Gas richness is a distance-independent quantity since both $\mhi$ and $\LV$ scale with distance squared. Therefore, a single spectrum allows us to meaningfully search for an HI reservoir in our targets anywhere within the wide bandpass.

The data were reduced using the standard GBTIDL\footnote{http://gbtidl.nrao.edu/} procedure $getps$.\ Before smoothing the raw spectra to our desired resolutions, we first removed narrow-band and broadband radio frequency interference (RFI). The latter is primarily present as a strong, intermittent GPS signal at $\nu\sim1.38 \, \mathrm{GHz}$ that is best removed by flagging the entire 5-second data dumps in which the RFI is present. On average, we flagged 20\% of the data to remove this feature. Narrow-band RFI presents itself throughout the entire spectrum as spurious, strong signals that span a few raw spectral channels.\ As scans and their constituent integrations are co-added and eventually smoothed, these narrow-band features may resemble the expected HI profile of a dwarf if not excised.\ To remove narrow-band RFI, we searched through all channels of each integration for signals that exceed 5 times the median absolute deviation.\ This threshold value was found to be the most effective at identifying RFI spikes while avoiding noise fluctuations. The values in these channels were then replaced with the median of the 200 surrounding channels.  

The calibrated, RFI-excised spectra were smoothed to multiple resolutions from $5-50 \, \kms$ and examined by eye to search for statistically significant emission. A representative RMS noise for each spectrum at $\Delta V = 25\,\kms$ resolution is given in column 11 of Table \ref{table:maintable}.\ We detect HI emission along the LOS to 5 targets (column 12 of Table \ref{table:maintable}). Their spectra are shown in Figure \ref{fig:detectspectra} at $\Delta V$ given in Table \ref{table:detectiontable}, which lists all other properties derived from these HI detections. We find no emission associated with the 22 remaining targets. These spectra, with velocity resolution, $\Delta V = 25\,\kms$, are shown in Figure \ref{fig:nondetectspectra}, and the corresponding upper limits on HI mass and gas richness are in Table \ref{table:upperlimits}.  

\floattable
\begin{deluxetable}{cccCCCcCcCCc}
\tablecaption{Target LSB Dwarf Candidate Properties \label{table:maintable}}

\tablehead{
\colhead{Name} & \colhead{RA} & \colhead{Dec} & \colhead{$m_g$} &
\colhead{$g-r$} & \colhead{$r_{eff}$} & \colhead{Ref} & \colhead{$D_{opt}$} & \colhead{Ref} & \colhead{Int. Time} &
\colhead{$\sigma_{25}$} & \colhead{HI}  \\
\colhead{} & \colhead{H:M:S} & \colhead{D:M:S} &  \colhead{(mag)} &
\colhead{(mag)} & \colhead{({arcsec})} &\colhead{} & \colhead{(Mpc)} & \colhead{} & \colhead{(hours)} & \colhead{(mJy)} & \colhead{Det?}\\
\colhead{(1)} & \colhead{(2)} & \colhead{(3)} & \colhead{(4)} & \colhead{(5)} &
\colhead{(6)} & \colhead{(7)} & \colhead{(8)} &
\colhead{(9)} & \colhead{(10)} & \colhead{(11)} & \colhead{(12)}
}
\startdata
DF-1 & 14:03:45.0 & 53:56:40& 19.4 \pm 0.1 & 0.60 \pm 0.14 &13.59 \pm 0.29&M16, B17& 6.37^{+0.35}_{-0.35} & D17 &1.4&0.49&  \\
DF-2 & 14:08:37.5 & 54:19:31 & 19.8 \pm 0.1 & 0.60 \pm 0.14 &10.35 \pm 0.68&M16, B17& 6.87^{+0.21}_{-0.30} & D17 &4.5&0.23&  \\
DF-3 & 14:03:05.7 & 53:36:56 & 20.3 \pm 0.1 & -0.10 \pm 0.14 &20.0 \pm 2.6&M16, B17& 6.52^{+0.25}_{-0.27} & D17 &1.3&0.39&  \\
DF-5 & 14:04:28.1 & 55:37:00 & 20.8 \pm 0.2 & 0.20 \pm 0.28 &10.8 \pm 2.6&M16, B17& > 17.5 & M16 &2.7&0.27&  Y\\
DwA* & 14:06:49.8& 53:44:29& 19.5 \pm 0.1 & 0.50 \pm 0.14 &10.92 \pm 0.23& K15, B17& 6.83^{+0.27}_{-0.26} & B19 &2.5&0.40&  \\
DwB & 14:08:43.1 & 55:09:57 & 20.8 \pm 0.1 & 0.80 \pm 0.22 &6.95 \pm 0.54& K15, B17& > 15.1 & B19 &2.9&0.42& Y \\
DwC & 14:05:18.0& 54:53:56& 20.5 \pm 0.2 & 0.70 \pm 0.28 &7.9 \pm 1.6& K15, B17& > 15.1 & B19 &5.5&0.21&  \\
DwD & 14:04:24.6& 53:16:19& 19.5 \pm 0.1 & 0.30 \pm 0.14 &9.16 \pm 0.47& K15, B17& > 15.1& B19 &1.9&0.49&  \\
Dw3 & 14:08:45.8 & 55:17:14 & 19.8 \pm 0.1 & 0.50 \pm 0.22 &7.11 \pm 0.42& B17, B17 & > 15.1 & B19 &1.7&0.39&  \\
Dw4 & 14:13:01.7 & 55:11:16 & 20.1 \pm 0.1 & 0.20 \pm 0.14 &6.88 \pm 0.33& B17, B17 & > 15.1 & B19 &3.6&0.25&  \\
Dw5 & 14:04:13.0 & 55:43:34 & 20.2 \pm 0.2 & 0.10 \pm 0.22 &7.64 \pm 0.86& B17, B17 & > 15.1 & B19 &5.7&0.22&  \\
Dw6 & 14:02:20.1 & 55:39:17 & 19.9 \pm 0.1 & 0.50 \pm 0.14 &8.34 \pm 0.37& B17, B17 & > 15.1 & B19 &1.6&0.38&  \\
Dw7 & 14:07:21.0 & 55:03:51 & 21.1 \pm 0.1 & 1.70 \pm 0.14 &4.7 \pm 0.20& B17, B17 & > 15.1 & B19 &4.4&0.29&  \\
Dw8 & 14:04:24.9 & 55:06:13 & 19.8 \pm 0.1 & 0.50 \pm 0.14 &5.70 \pm 0.20& B17, B17 & > 15.1 & B19 &1.6&0.34&  \\
Dw13 & 14:08:01.2 & 54:22:30 & 20.4 \pm 0.1 & 0.60 \pm 0.14 &3.91 \pm 0.1& B17, B17 & > 15.1 & B19 &3.6&0.29&  \\
Dw19 & 14:10:20.1 & 54:45:50 & 20.4 \pm 0.1 & 0.60 \pm 0.14 &4.56 \pm 0.23& B17, B17 & > 10.2 & C19 &5.4&0.35&  \\
Dw26 & 14:08:50.4 & 53:27:24 & 20.2 \pm 0.1 & 0.40 \pm 0.14 &5.08 \pm 0.10& B17, B17 & - & - &3.4&0.29& Y \\
Dw31 & 14:07:41.7 & 54:35:18 & 19.4 \pm 0.1 & 0.60 \pm 0.14 &5.91 \pm 0.23& B17, B17 & > 10.8 & C19 &1.0&0.61&  \\
Dw32 & 14:07:46.4 & 54:15:26 & 19.0 \pm 0.1 & 0.20 \pm 0.14 &10.70 \pm 0.25& B17, B17 & > 12.6 & C19 &1.0&0.76&  \\
Dw33 & 14:08:33.8 & 55:26:49 & 19.8 \pm 0.1 & 0.60 \pm 0.22 &4.59 \pm 0.21& B17, B17 & 18.6^{+3.4}_{-2.9} & C19 &2.0&0.47&  \\
Dw38 & 14:01:17.6 & 54:21:14 & 20.1 \pm 0.1 & 0.30 \pm 0.14 &4.33 \pm 0.16& B17, B17 & > 26.3 & C19 &3.3&0.62&  \\
dw1343+58 & 13:43:07 & 58:13:40 & 15.54 \pm 0.30 & 0.37 \pm 0.42  &28.6 \pm 1.3& M17, M17 & - & - &0.2&2.35& Y \\
dw1355+51 & 13:55:11 & 51:54:29 & 18.76 \pm 0.30 & 0.67 \pm 0.42 &7.44 \pm 1.3& M17, M17 & - & - &0.5&0.79&  \\
dw1408+56 & 14:08:41 & 56:55:38 & 18.01 \pm 0.30 & 0.51 \pm 0.42 &11.0 \pm 1.3& M17, M17 & 11.2^{+3.4}_{-2.4} & C19 &0.2&0.52& Y \\
dw1412+56 & 14:12:11 & 56:08:31 & 19.46 \pm 0.30 & 0.71 \pm 0.42 &8.08 \pm 1.3& M17, M17 & > 9.0 & C19 &0.7&0.52&  \\
dw1416+57 & 14:16:59 & 57:54:39 & 19.06 \pm 0.30 & 0.23 \pm 0.42 &7.35 \pm 1.3& M17, M17 & - & - &0.8&0.66&  \\
dw1446+58 & 14:47:00 & 58:34:04 & 18.46 \pm 0.30 & 0.56 \pm 0.42 &8.97 \pm 1.3& M17, M17 & - & - &0.3&0.85&  \\
\enddata
\tablecomments{col.(1): Adopted LSB dwarf candidate name. cols.(2) and (3): J2000 position of optical centroid, which corresponds to our GBT LOS. cols.(4) and (5): $g-$band apparent magnitudes and $g-r$ colours. col.(6): Effective radius of LSB dwarf candidate. col.(7): First reference lists origin of candidate name and position in cols.(1)-(3), second lists source of photometric measurements in cols. (4)-(6). col.(8):  Adopted distance constraint from reference in col.(9). col.(10): Total effective GBT integration time, including the ON+OFF positions and subtracting any time lost due to RFI flagging. col.(11): Representative RMS noise of the spectrum at a velocity resolution of $\Delta V = 25 \, \kms$. col.(12): Flag indicating whether or not we detect HI emission associated with dwarf candidate. References: M14 = \citet{2014MerritLSBsM101}, K15 = \citet{2015AstBu..70..379K}, M16 = \citet{2016ApJ...833..168M}, B17 = \citet{Paulspaper}, D17 = \citet{2017M101LuminosityFn}, M17 = \citet{Mullerspaper}, B19 = \citet{Paulm101hst}, C19 = \citet{CarlstenSBFM101}. * - Target initially reported by K15 but GBT LOS corresponds to position from \citet{JavanmardiDGSAT}; given the 9.1 arcminute GBT beam, an offset of a few arcseconds is negligible.}
\end{deluxetable}

\section{Results} \label{sec:results}
\subsection{ HI Detections} \label{subsec:detections}
\begin{figure*}[htb!]
\includegraphics[width=18cm]{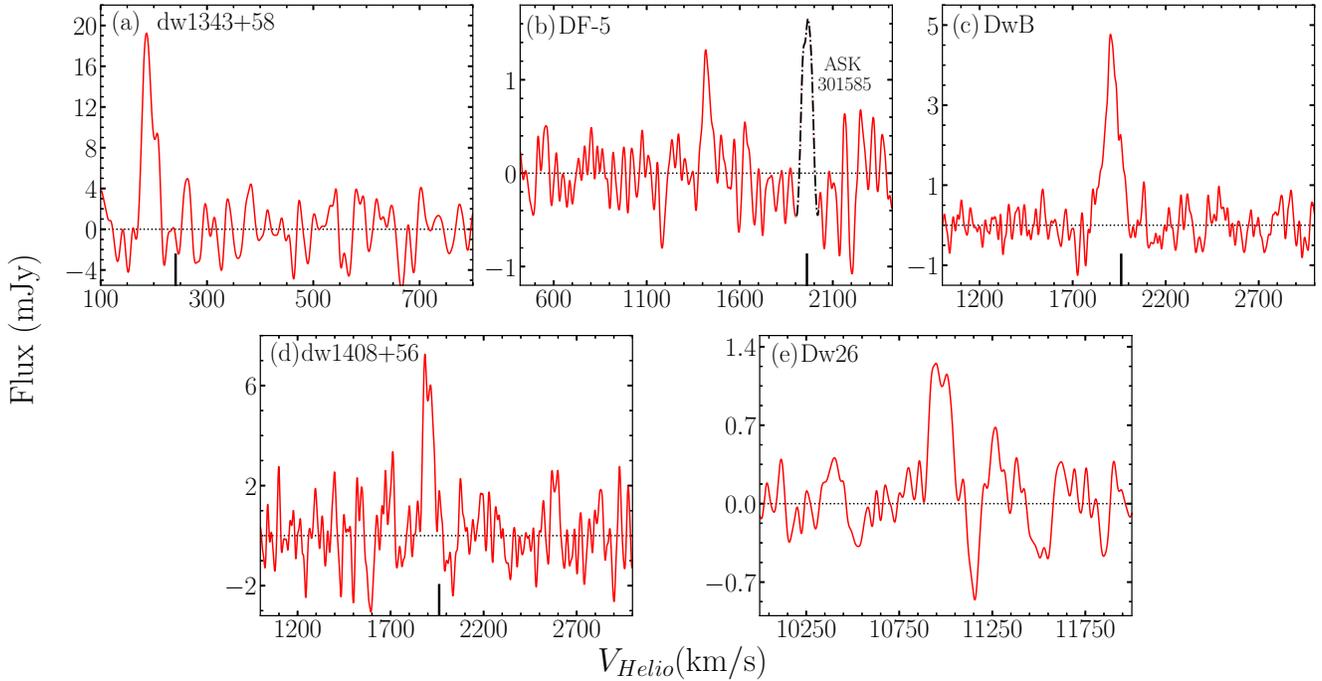}
\caption{ HI detections along the LOS to LSB dwarf candidates in the region around M101. Target names are in the top-left corner of each panel and the spectral resolutions $\Delta V$ of the plotted spectra are in Table \ref{table:detectiontable}. The spectrum in (a) is off-center as Milky Way emission dominates at lower velocities. The dash-dotted feature in (b) is the HI emission associated with ASK 301585 (see text). The vertical, black lines below the spectra indicate the $\vsys$ of M101 for (a) and the NGC~5485 group for (b), (c), and (d). The derived properties of the HI detections are provided in Table \ref{table:detectiontable}.}
\label{fig:detectspectra}
\end{figure*}

We detect HI along the LOS to 5 of the LSB dwarf candidates in our sample, and their spectra are shown in Figure \ref{fig:detectspectra}. At our observing frequency of $\sim$ 1.4 GHz, the full-width half maximum of the GBT beam, $\sim \mathrm{9.1'}$, encompasses the entire stellar component of our dwarf targets. The GBT beam response is well understood down to $\approx -30 \mathrm{dB}$ \citep[e.g.][]{GBTbeam}, and we can assess the extent to which gas-rich sources near the LOS contaminate our spectra. For all of our targets, we search through NED\footnote{The NASA/IPAC Extragalactic Database (NED) is operated by the Jet Propulsion Laboratory, California Institute of Technology, under contract with the National Aeronautics and Space Administration.} and SDSS imaging catalogs \citep{SDSSDR15}
for objects within $30'$ of the LOS that may present themselves as gas-rich interlopers in our spectra (the spectrum of one interloper, ASK 301585, is visible in panel (b) of Fig.\ \ref{fig:detectspectra}).\ We find no such interlopers for our HI detections, and conclude that they are the HI counterparts to the LSB dwarf candidates along the LOS. The HI properties that we derive for the detections are given in Table \ref{table:detectiontable}.

We first derive distance-independent quantities, systemic velocities ($\vsys$) and velocity widths ($\wfty$), following the methods of \citet{2005ApJS..160..149S}.\ We fit first-order polynomials to both edges of each HI profile in Figure \ref{fig:detectspectra} between 15\% and 85\% of the peak flux value.\ To determine $\vsys$ and $\wfty$, we first determine the velocities corresponding to the 50\% flux level for each polynomial fit. The mean of these two values provides $\vsys$, while the difference provides $\wfty$ which is also corrected for instrumental and cosmological redshift broadening to produce $W_{50,c}$.\ We do not correct for the effect of inclination on $\wfty$. The uncertainties on the instrumental broadening correction, which we take to be 50\%, dominate those on $\vsys$ and $\wfty$ (see \citealt{2005ApJS..160..149S}).  
 
We estimate kinematic distances, $D_{HI}$, for DF-5, DwB, dw1408+56, and Dw26 using $\vsys$ of our detections and assuming $H_{0}=70 \, \kms \mathrm{Mpc}^{-1}$. The value of $D_{HI}$ for DF-5 is consistent with the lower limits on $D_{opt}$ from M16 and C19.\ The values of $\vsys$ that we derive for DwB and dw1408+56 are similar to those of the background group containing NGC~5485, implying an association.\ The SBF lower limit on $D_{opt}$ from C19 for DwB is consistent with our distance estimate, while their $D_{opt}$ estimate for dw1408+56 is not.\ C19 note that their distance for this latter object may be unreliable due to its unusual morphology, and they conclude that it likely lies in the background of M101 as we find here. Our measurement of $D_{HI}$ for Dw26 is the first distance estimate for this system (see also B17). For the purposes of this work, we assign a distance of $D_{HI} = D_{M101} = 7.0 \, \mathrm{Mpc}$ to dw1343+58 because of its similar recessional velocity to the broader M101 group complex.

We calculate the HI flux, $S_{HI}={\int}S{\delta V}$, by integrating over the line profile.\ HI masses, $\mhi$, are determined using the standard equation for an optically thin gas \citep{1984AJ.....89..758H}: 
\begin{equation}M_{HI}=2.356\times 10^{5}(D_{HI})^{2}S_{HI} \, \mathrm{\msun},\end{equation} 
where $D_{HI}$ is in Mpc and $S_{HI}$ is in Jy $\kms$. Uncertainties are determined following the methods of \cite{2005ApJS..160..149S}.\ Using $m_{g}$ and $g-r$ from Table \ref{table:maintable}, the relations of \citet{Jestervband}, and $D_{HI}$ we estimate the $V-$band luminosities, $\LV$, and the gas richnesses, $\mhi/\LV$.\  These values are tabulated in Table \ref{table:detectiontable}.

\begin{deluxetable*}{cCCCCCCCCC}[htb!]
\caption{Properties of LSB Dwarf Candidates with HI detections}
\label{table:detectiontable}
\tablehead{
\colhead{Name} & \colhead{$\Delta V$} & \colhead{$\sigma_{\Delta V}$} & \colhead{$V_{sys}$} & \colhead{$W_{50,c}$} & \colhead{$S_{HI}$} & \colhead{$D_{HI}$} & \colhead{log($\LV$)} &
\colhead{log($\mhi$)} & \colhead{$(\frac{\mhi}{\LV})$}  \\
\colhead{} & \colhead{($\kms$)} & \colhead{(mJy)} & \colhead{($\kms$)} & \colhead{($\kms$)} & \colhead{(Jy$\,\kms$)} & \colhead{(Mpc)} &
\colhead{(log[$\lsun$])} & 
\colhead{(log[$\msun$])} & \colhead{$(\frac{\msun}{\lsun})$} \\
\colhead{(1)} & \colhead{(2)} & \colhead{(3)} & \colhead{(4)} & \colhead{(5)} &
\colhead{(6)} & \colhead{(7)} & \colhead{(8)} &
\colhead{(9)} & \colhead{(10)}
}

\startdata
dw1343+58 & 10 & 2.3 & 195 \pm 1 & 37 \pm 1 & 0.62 \pm 0.12 & 7.0^{*} & 7.48 \pm 0.12 & 6.85 \pm 0.09 & 0.23 \pm 0.08 \\
DF5 & 20 & 0.3 & 1424 \pm 4 & 36 \pm 6 & 0.06 \pm 0.02 & 20.3 & 6.26 \pm 0.08 & 6.79 \pm 0.18 & 3.4 \pm 1.5 \\
DwB & 15 & 0.4 & 1913 \pm 4 & 69 \pm 5 & 0.37 \pm 0.04 & 27.3 & 6.65 \pm 0.04 & 7.81 \pm 0.05 & 14 \pm 2 \\
dw1408+56 & 15 &  1.0 & 1904 \pm 2 & 55 \pm 4 & 0.38 \pm 0.08 & 27.2 & 7.70 \pm 0.12 & 7.82 \pm 0.10 & 1.3 \pm 0.4 \\ 
Dw26 & 30 & 0.2 & 10972 \pm 12 & 81 \pm 16 & 0.15 \pm 0.04 & 156 & 8.32 \pm 0.04 & 8.94 \pm 0.12 & 4.2 \pm 1.3 \\
\enddata
\tablecomments{col.(2): Velocity resolution of spectrum used to compute HI properties (see Figure \ref{fig:detectspectra}). col.(3): RMS noise of spectrum at $\Delta V$ in col.(2). col.(4): Heliocentric systemic velocity. col.(5): Velocity width of the HI detection, corrected for cosmological redshift and instrumental broadening. col.(6): Integrated HI flux. col.(7): Kinematic distance estimated using the Hubble-Lema\^{i}tre Law, $\vsys$ and $\mathrm{H}_0 = 70 \, \kms\,\mathrm{Mpc}^{-1}$, except for dw1343+58(*) for which the M101 distance is adopted; see text for explanation. col.(8): Logarithm of V-band luminosity calculated using $m_g$ and $g-r$ in Table \ref{table:maintable} and $D_{HI}$ from col.(7). col.(9): Logarithm of HI mass calculated from Eq.(1) using $S_{HI}$ in col.(6) and $D_{HI}$ in col.(7). col.(10): HI-mass to V-band luminosity ratio (gas richness).}
\end{deluxetable*}

\subsection{ HI Non-detections} \label{subsec:nondetections}
\begin{figure*}
\includegraphics[width=18cm]{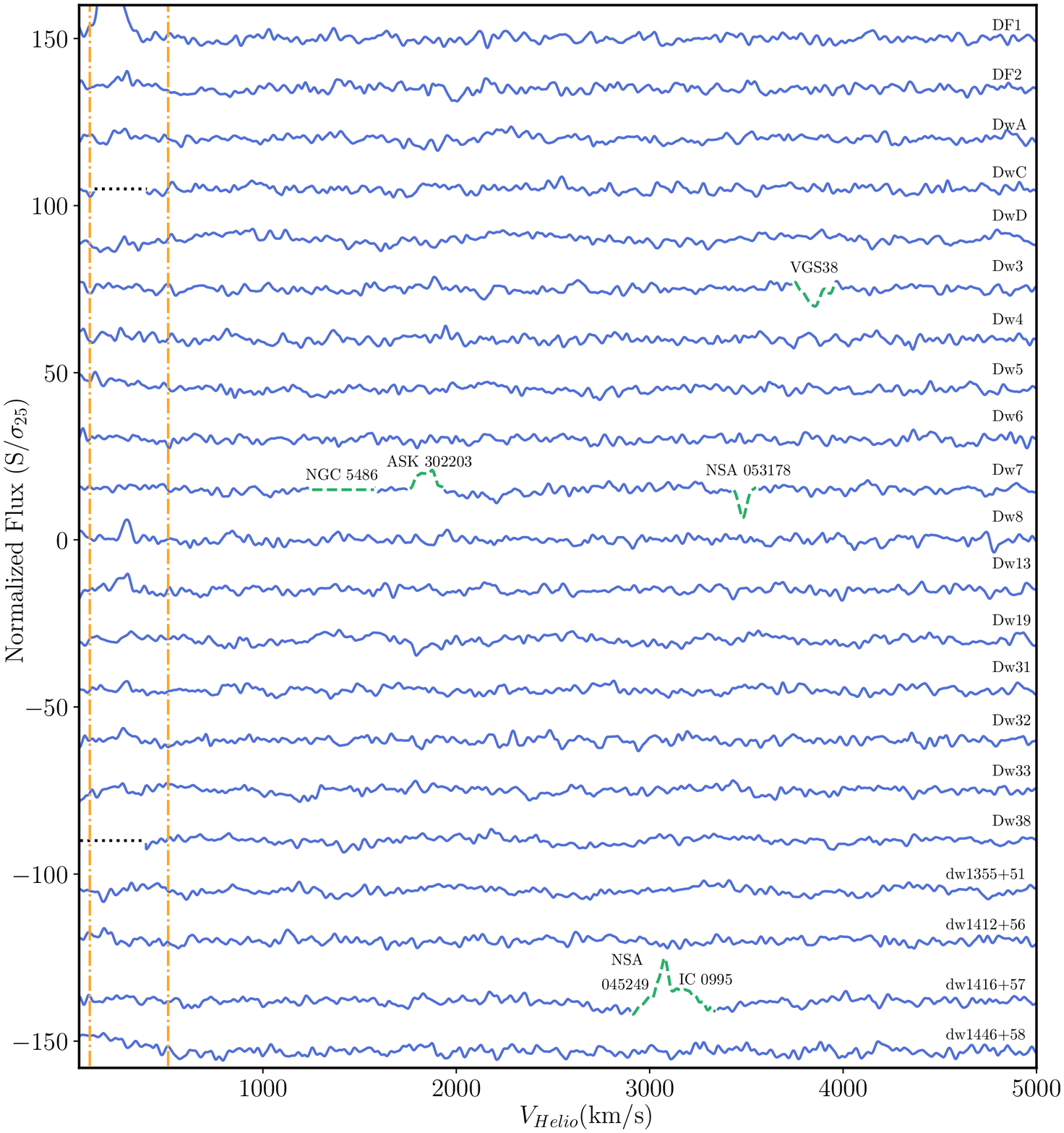}
\caption{Spectra along the LOS to LSB dwarf candidates for which no associated HI emission was found, at a resolution of $\Delta V = 25 \, \kms$ and normalized by the RMS values listed in Table \ref{table:maintable}. The spectra are presented in the same order as in Table \ref{table:upperlimits} but with DF-3 omitted because the spectral setup for that target was different; see text for explanation. Target names are shown above their respective spectrum at the right side of the figure. The vertical offset between the spectra is arbitrary to better display them. Parts of the spectra where an interloping HI signal of a nearby gas-rich galaxy is present in the ON or OFF scans are shown as green dashed lines. The vertical orange lines indicate the velocity region within which HI emission from M101 itself may be present. Two spectra (DwC and Dw38) have had M101 emission removed and replaced with black dotted lines for aesthetic purposes.}
\label{fig:nondetectspectra}
\end{figure*}

We find no HI signal along the LOS to the remaining 22 LSB dwarf candidates in our sample that we can attribute to these objects.\ We show their spectra in Figure \ref{fig:nondetectspectra}, normalized by the RMS values $\sigma_{25}$ listed in Table \ref{table:maintable}.\ We search through NED and SDSS imaging catalogs for potential interlopers within $30'$ of our targets.\ In Figure \ref{fig:nondetectspectra}, HI emission from nearby objects in the spectra are shown by green dashed lines with object names given above them.\ We note that the features are positive if the corresponding object is located in the ON, and negative if they are located in an OFF.

We use $\sigma_{25}$ to place single, $25\,\kms$-channel 5$\sigma$ upper limits on the HI mass of the LSB dwarf candidate non-detections, $\mhilim$, using a modified version of Eq.(1),  
\begin{equation} \mhilim=2.945\times 10^{7}D^{2}({\sigma_{25}})\, \mathrm{\msun},\end{equation}
where $\sigma_{25}$ is in $\kms$ and $D$ is in Mpc.\ For objects with $D_{opt}$ estimates or lower limits in Table \ref{table:maintable}, we use that value as $D$ in Eq.(2).\ For objects lacking a $D_{opt}$ we compute $\mhilim$  by setting $D$ to the M101 distance,  $D_{M101} = 7.0 $ Mpc.\ We also estimate the HI mass of these objects if they are located at the distance of the NGC~5485 group, $(\mhilim)_{BG}$, by setting $D=D_{BG}=27\,\mathrm{Mpc}$ in Eq.(2).\ Both $\mhilim$ and $(\mhilim)_{BG}$ are given in Table \ref{table:upperlimits}, along with the corresponding (distance-independent) upper limit $\mhilim/\LV$ on gas richness.

\subsection{Optical and Gas Properties} \label{subsec:optHI}
\begin{figure*}[h!]
\includegraphics[width=18cm]{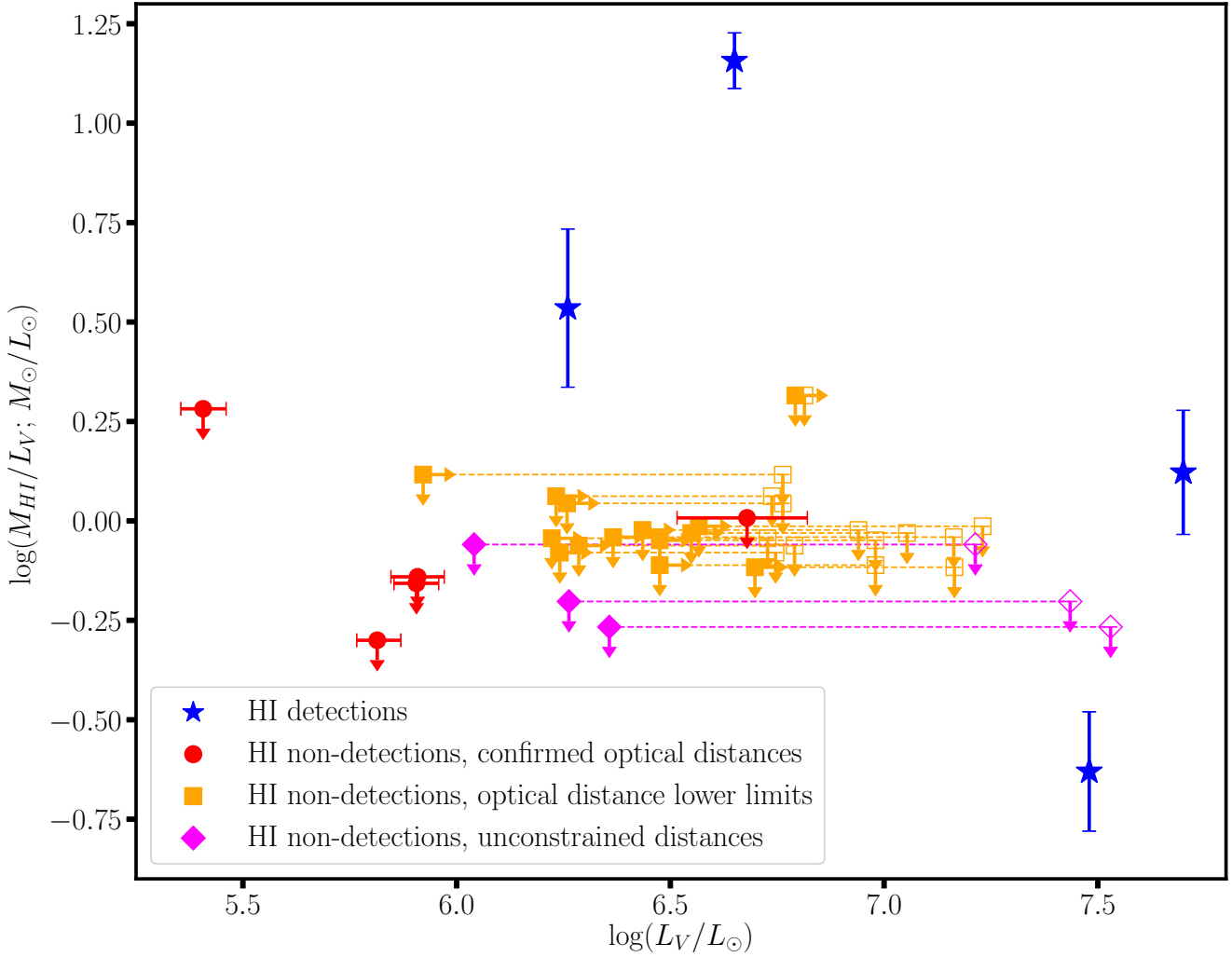}
\caption{$\mhi/\LV$ (blue stars) and $\mhilim/\LV$ (red, orange, and magenta downward arrows) as a function of $\LV$ for LSB dwarf candidates in our sample. Among HI non-detections, red circles show objects with measured $D_{opt}$, orange squares show objects with lower limits on $D_{opt}$, and magenta diamonds show objects with no distance constraints. We place orange squares at $\LV$ corresponding to the $D_{opt}$ lower limits (filled symbol and rightward arrows) and $D_{BG}$ (empty symbol) of the NGC~5485 group for the corresponding objects, and magenta diamonds at $\LV$ corresponding to $D_{M101}$ (filled) and $D_{BG}$ (empty). Pairs of symbols for the same objects are connected by horizontal dashed lines. We omit one gas-rich target for aesthetic purposes, Dw26 (log${(\mhi/\LV)}$ = 0.62, log$(\LV)$ = 8.32).}
\label{fig:gasrichness-lum}
\end{figure*}

With the derived HI properties for our sample in-hand we can make comparisons to their optical properties. Figure \ref{fig:gasrichness-lum} plots ${\mhi/\LV}$ for detections and $(\mhilim)/\LV$ for non-detections as a function of $\LV$.\ We separate our sample into four groups: objects with HI detections (blue stars), non-detections with confirmed distances, $D_{opt}$ (red circles), non-detections with lower limits on $D_{opt}$ (orange squares), and non-detections with no distance measure (magenta diamonds). Since $\mhi/\LV$ is distance independent, the location of the points along the y-axis is known for all of the points.\ For non-detections with lower limits on $D_{opt}$ or no available estimates, we place symbols at both the $D_{opt}$ lower limit and $D_{BG}$ (filled and empty orange squares) or $D_{M101}$ and $D_{BG}$ (filled and empty magenta diamonds) along the x-axis and connect the two with a horizontal dashed line. For clarity, we do not include HI detection Dw26, with $\mathrm{log}(\LV;\lsun)=8.3$. Considering this object in addition to those plotted in Fig.\ \ref{fig:gasrichness-lum}, over half of the HI detections correspond to objects with higher $\LV$ than the non-detections even if the latter are placed at the distance of the NGC~5485 group. 

\begin{figure*}[h!]
\includegraphics[width=18cm]{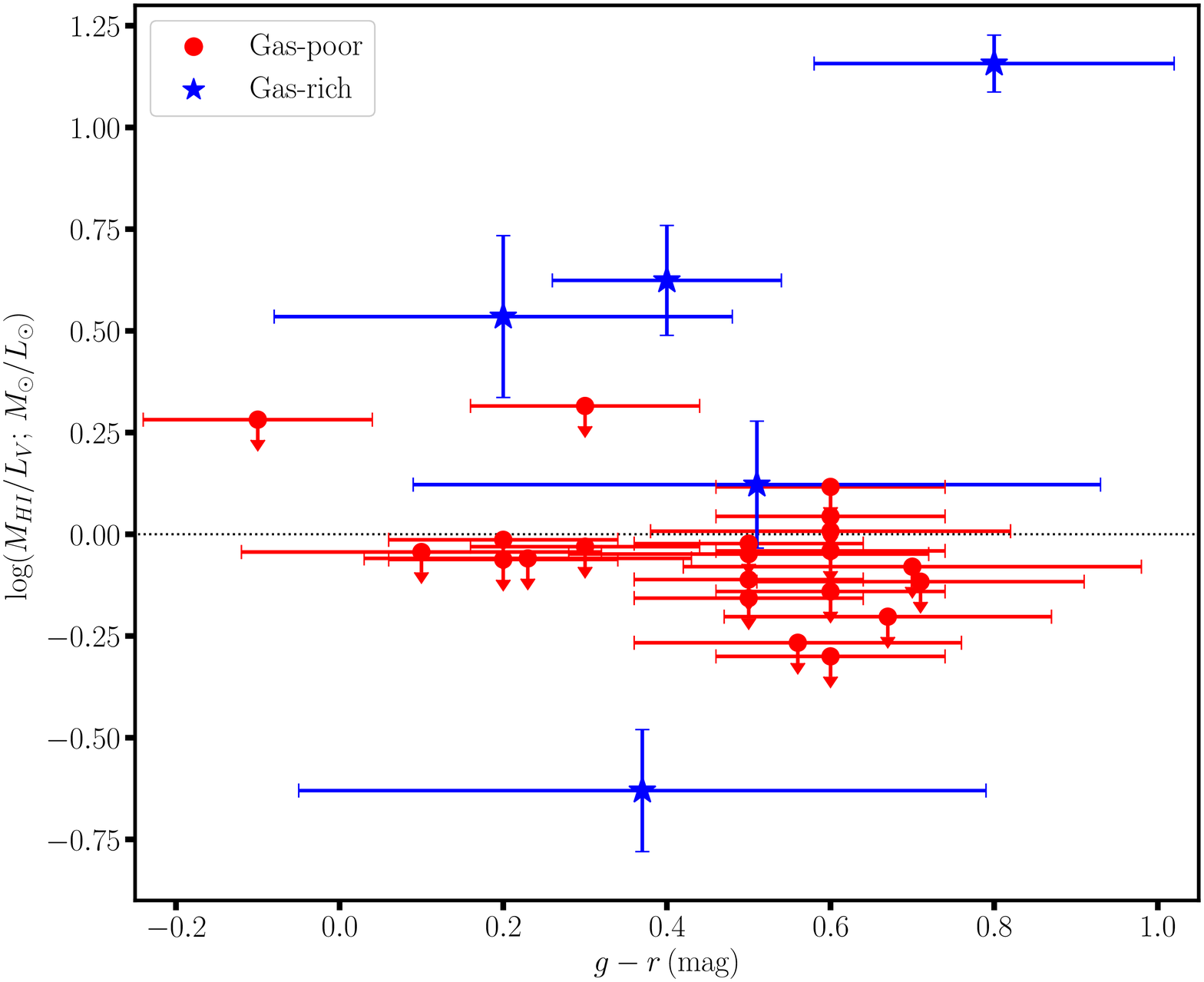}
\caption{$\mhi/\LV$ (blue stars) and $\mhilim/\LV$ (red circles with downward arrows) versus $g-r$ colour for our sample. The horizontal dotted line shows ${\mhi/\LV=1\msun/\lsun}$, typical for dwarf galaxies \citep[e.g.][]{huangalfalafa,Bradford2015MAIN}. We omit one gas-poor outlier in colour for aesthetic purposes, Dw7 (log${(\mhilim/\LV)}$ = 0.05, $g-r$ = 1.7).}
\label{fig:gasrichness-colour}
\end{figure*}

Figure \ref{fig:gasrichness-colour} shows  gas richness versus $g-r$ for the sample, separated into HI detections (blue stars) and HI non-detections (red circles with downward arrows). The horizontal dotted line indicates $\mhi/\LV = 1 \msun/\lsun$ which is typical for dwarf galaxies \citep[e.g.][]{huangalfalafa,Bradford2015MAIN}. Both our detections and non-detections have a wide range of colours. We discuss this further in the following section.

\begin{deluxetable}{cCCC}
\caption{ HI Upper Limits for Non-detections}
\label{table:upperlimits}
\tablehead{
\colhead{Name} & \colhead{log$(\mhilim)$} & \colhead{log$(\mhilim)_{BG}$} & \colhead{$({\mhilim}/{\LV})$}  \\
\colhead{} & \colhead{(log[$\msun$])}& \colhead{(log[$\msun$])} & \colhead{$({\msun}/{\lsun})$} \\
\colhead{(1)} & \colhead{(2)} & \colhead{(3)} & \colhead{(4)}
}
\startdata
DF-1 &5.77&-&0.70 \\
DF-2 &5.51&-&0.49 \\ 
DF-3 &5.69&-&1.86 \\
DwA &5.75&-&0.68 \\
DwC &6.16&6.67&0.81 \\
DwD &6.52&7.02&0.91 \\
Dw3 &6.43&6.93&0.87 \\
Dw4 &6.22&6.73&0.84 \\
Dw5 &6.18&6.68&0.88 \\
Dw6 &6.41&6.92&0.92 \\
Dw7 &6.29&6.80&1.12 \\
Dw8&6.36&6.87&0.75 \\
Dw13&6.30&6.81&1.08 \\ 
Dw19&6.04&6.88&1.27 \\ 
Dw31&6.33&7.12&0.89 \\
Dw32&6.55&7.22&0.94 \\
Dw33&6.69&7.00&0.99 \\
Dw38&7.11&7.13&2.01 \\
dw1355+51&6.06&7.23&0.61 \\
dw1412+56&6.58&7.05&0.74 \\ 
dw1416+57&5.98&7.15&0.85 \\
dw1446+58&6.09&7.26&0.53 \\
\enddata
\tablecomments{col.(2): $5\sigma$ upper limit on $\mhi$ calculated from Eq.(2) using $D=D_{opt}$ and $\sigma_{25}$ from Table \ref{table:maintable}. col.(3): Same as col.(2) but instead $D=D_{BG} = 27 \, \mathrm{Mpc}$ of the NGC~5485 group for systems without $D_{opt}$ estimates or with lower limits. An exception is Dw33 which has a $D_{opt}$ estimate, however, we suggest it may reside in the background group (see Section \ref{subsec:gaspoor}).col.(4): Upper limit on the gas richness (which is distance independent).}
\end{deluxetable}

\section{Discussion} \label{sec:Discussion}
With the gas properties of our HI detections and non-detections in hand, we consider their most likely physical associations along with their implications for the gas richnesses of satellite systems on both the galaxy and group scales.

\subsection{Associations of HI Detections} \label{subsec:gasrich}
The systemic velocities of our 5 detections allow us to determine whether they are associated with M101 or the NGC~5485 group in the background (see Section \ref{sec:intro}), or whether they are at some other location along the LOS.

The systemic velocity of the HI detection of dw1343+58, $V_{sys}=195 \,\kms$ (Table \ref{table:detectiontable}), confirms that it is a nearby dwarf. Its projected separation from M101, $d_{proj}\sim 580 \mathrm{kpc}$, and similar $\vsys$ to M101 ($V_{M101} = 241 \, \kms$) suggests that it is likely a member of the M101 group. The SDSS spectrum for dw1343+58, $V_{SDSS}= 165 \pm 33 \,\kms$, is consistent with the $\vsys$ that we derive, although the SDSS pipeline mis-classified it as a star. Given its optical morphology, M17 suggest that dw1343+58 is a blue compact dwarf (BCD). The relatively low gas richness of dw1343+58 compared to typical star-forming dwarfs \citep[e.g.][]{huangalfalafa,Bradford2015MAIN}, and the UV counterpart that we identify in our search of archival GALEX data provide further evidence in support of this classification \citep{BCDsgeneral,bcdgil}.

On the other hand, DwB and dw1408+56 have $\vsys$ consistent with that of the NGC~5485 group.\ Using their kinematic distances and projected separations, we estimate that the physical separations of DwB and dw1408+56 from NGC~5485 are $\sim 450$ kpc and $\sim 940$ kpc, respectively.\ Given its kinematic distance, the properties of dw1408+56 resemble those of a gas-rich ultra-diffuse galaxy \citep[UDG;\,$R_{eff} > 1.5 \, \mathrm{kpc}, \mu_0 > 24 \,\mathrm{mag\,arcsec^{-2}}$;][]{UDGsvandokkum} progenitor with $R_{eff} = 1.45 \pm 0.17$ kpc and $\mu_{r,0} = 23.28 \pm 0.06 \,\mathrm{mag\,arcsec^{-2}}$ \citep{HCGUDGs}. This finding is broadly consistent with UDG number-halo mass relation estimates \citep[e.g.][]{VanderburgUDGs}, which imply that the number of UDGs in a group like that containing NGC~5485 should be of order 1.

DF-5 was suggested to be a member of the NGC~5485 group since HST imaging does not resolve any stars (M16), and our HI detection confirms that it is in the background of M101 with $V_{sys}=1424 \kms$. While this $\vsys$ differs by $\sim \,500 \kms$ from that of the group \citep{Galaxygroups2massTully}, the small projected separation between DF-5 and NGC~5485 implies that a physical association is nonetheless feasible.\ We return to this issue in Section \ref{subsec:gaspoor}.

Dw26 is our most distant HI detection with $D_{HI} = 156 \,\mathrm{Mpc}$.\ At such a distance, Dw26 is a star-forming galaxy well in the background of both M101 and the NGC~5485 group. The vast range of distances that we derive from our HI detections in the M101 region demonstrates the need for follow up observations of optical LSB features to determine their physical properties and associations.

The relationship between stellar mass, colour and gas richness in the high surface brightness galaxy population is well-studied \citep{GASS,huangalfalafa,Bradford2015MAIN,Brown2015}: at fixed stellar mass, bluer systems are more gas-rich. By contrast, Fig.\ \ref{fig:gasrichness-colour} illustrates that our HI detections exhibit the same broad range in $g-r$ as our non-detections, and therefore that the reported colours do not predict gas richness.\ While it is possible that star formation is regulated differently in these faint systems compared to the broader galaxy population \citep{Wheeler2015,FIRE-lowmass,NIHAOLSB}, we deem it more likely that the large photometric uncertainties at such low surface brightness (B17, M17), evidenced by the sizeable error bars in Fig.\ \ref{fig:gasrichness-colour}, preclude the use of colour as an indicator of the presence of gas. This further underscores the utility of follow-up HI observations to constrain the gas content and star formation activity in the LSB galaxy population. 

\subsection{ HI non-detections and the NGC~5485 group} \label{subsec:gaspoor}

The majority of our follow-up HI observations resulted in non-detections, implying that the LSB dwarf candidates along those LOS are gas poor.\ Among them, 15/22 have distance constraints or lower limits from optical follow-up that place them in the background of M101 (see Table \ref{table:maintable}).\  Because field galaxies are almost universally star forming and HI rich \citep{GehaNSAfielddwarfs,Bradford2015MAIN} whereas our non-detections are clearly gas poor, we consider the likelihood that all of our non-detections are associated with the NGC~5485 background group and have been stripped of their HI reservoirs through environmental processes \citep[e.g.][]{RPS-Gatto2013,Wetzel2015,RPS-SimpsonAuriga}.

\begin{figure*}[h!]
\includegraphics[width=18cm]{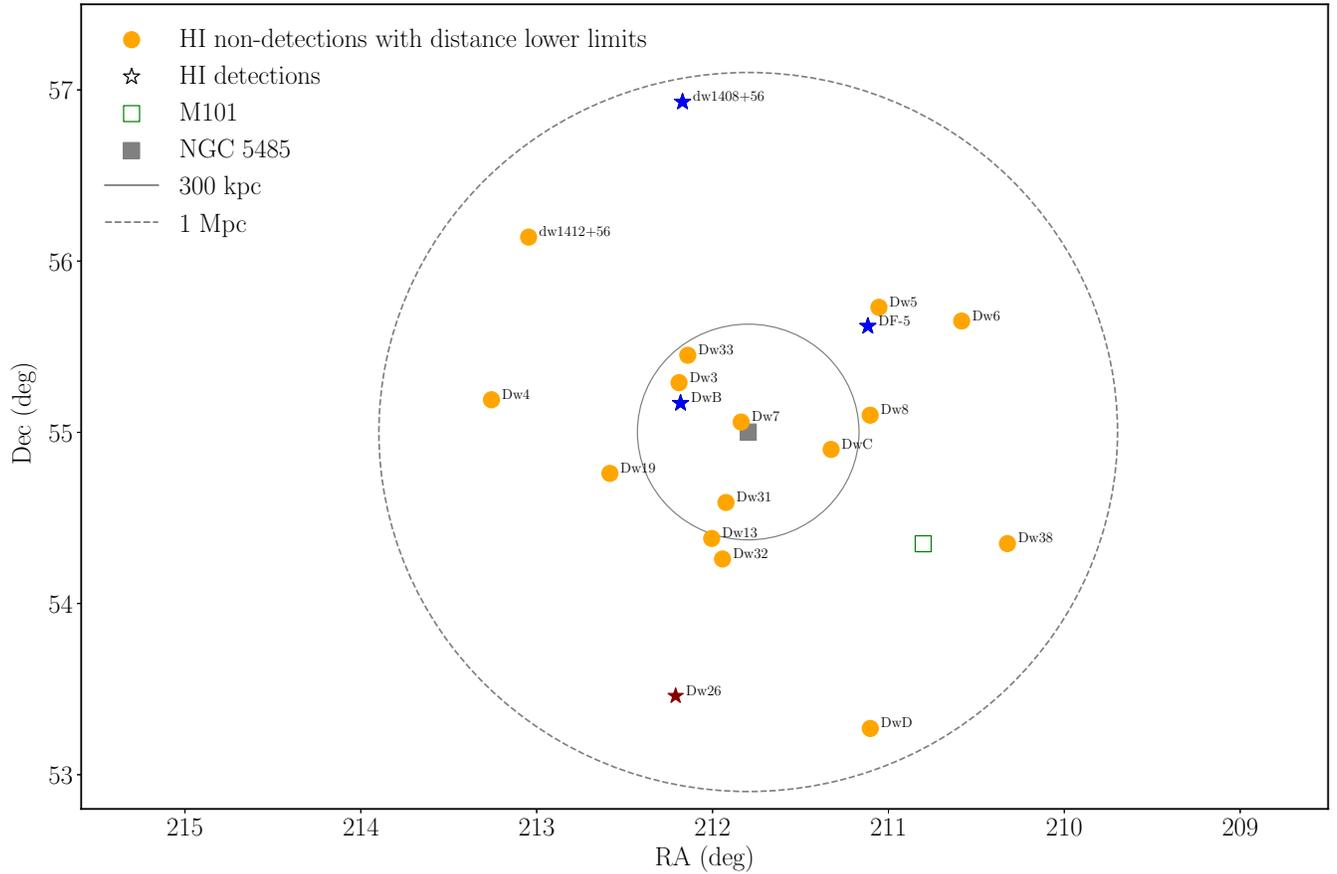}
\caption{Targets in our sample that project within $\sim 2$ deg ($=1$ Mpc at the NGC~5485 distance; dashed circle) of NGC~5485 (solid grey square) and that have no known association with M101. Objects confirmed to be in the background of M101 via our HI detections are shown as stars: filled in blue are those associated with the NGC 5485 group and the dark red star is Dw26 in the distant background. Objects that are undetected in HI but have optical distance measurements or lower limits that place them beyond M101 are shown as yellow circles. For reference, the open square shows the location of M101 in the foreground. The open grey circle centered on NGC~5485 has a radius of 300 kpc at its distance, representative of literature estimates for the virial radius of the corresponding group \citep{Galaxygroups2massTully,SaulderNGC5485,karaandmakaM101}.}
\label{fig:m101dist}
\end{figure*}

Figure \ref{fig:m101dist} shows all of the targets in our sample that have no known association with M101 and project within $\sim 2$ degrees of NGC~5485 ($=1$ Mpc at $D_{BG}$; dashed circle). These targets are either confirmed to be in the background of M101 via our HI detections (stars), are undetected in HI but have optical distance measurements or lower limits that place them beyond M101 (circles). Fig. \ref{fig:m101dist} suggests that, if our HI non-detections are at the NGC~5485 group distance and have been stripped of their gas due to environmental effects, then the sphere of influence of that group has a radius of at least $\sim 1$ Mpc. Observations show that the fraction of gas-rich or star-forming satellites depends on group mass \citep{2017browngasstripping, 2019schaefer}, while simulations suggest that the environmental influence of a group extends beyond its virial radius $R_{vir}$ \citep{2013bahe,2014cengasloss}. The latter suggest that the HI non-detections in Fig. \ref{fig:m101dist} could plausibly stem from gas stripping within $\sim2-3\,R_{vir}$ of the NGC~5485 group, which is broadly consistent with literature estimates of its $R_{vir}$ \citep[solid circle in Fig. \ref{fig:m101dist}, ][]{Galaxygroups2massTully,SaulderNGC5485,karaandmakaM101}.  

Indeed, the properties of both our detections and non-detections are consistent with a scenario in which almost all are physically associated with the NGC~5485 group. In this picture, the non-detections correspond to galaxies that have had their gas stripped by the NGC~5485 intra-group medium or by tides, as the gas in satellite galaxies is not expected to survive after first infall \citep{Wetzel2015}.  The HI detections correspond to galaxies on the group outskirts that have not lost their gas or have yet to be processed by the group environment. NGC~5485 has stellar stream-like features which are likely due to past interactions \citep{karaandmakaM101}, further supporting this scenario.

We consider each of the HI detections in turn to assess whether or not their properties are consistent with this scenario. The large projected separation between dw1408+56 and the group center as well as their correspondence in  $\vsys$ suggest that this object is indeed on the group outskirts. Similarly, the large difference in $\vsys$ between DF-5 and the NGC~5485 group ($\sim 500 \kms$) suggests that it is on its first infall with a large peculiar velocity. By contrast, DwB projects near the group center and has a similar $\vsys$; if it is indeed near the group center it should be gas-poor. This can be reconciled with our scenario if DwB is separated from the NGC~5485 group along the line-of-sight. In this case, the correspondence between the NGC~5485 group $\vsys$ and that of DwB would stem from the peculiar velocity of the latter because it is on first infall.

We note that the only HI follow-up target plotted in Fig.\ \ref{fig:m101dist} that is known not to be associated\footnote {We note that the C19 SBF distance for Dw33 places it in the background of M101 but also in the foreground of NGC~5485. However, since C19 caution that the uncertainties on their Dw33 distance measurement may be underestimated, we deem it plausible that Dw33 is in fact associated with NGC~5485.} with the NGC~5485 group is Dw26, which our HI detection places far in the background ($D\sim 156\,$Mpc). However, it is unlikely that a significant number of our non-detections are at such large distances: a mechanism for stripping them in a lower density environment needs to be found. We therefore posit that the most plausible origin for the majority of our non-detections is that they belong to the NGC~5485 group.

\subsection{Satellite Gas Richness}
\begin{figure*}[h!]
\includegraphics[width=18cm]{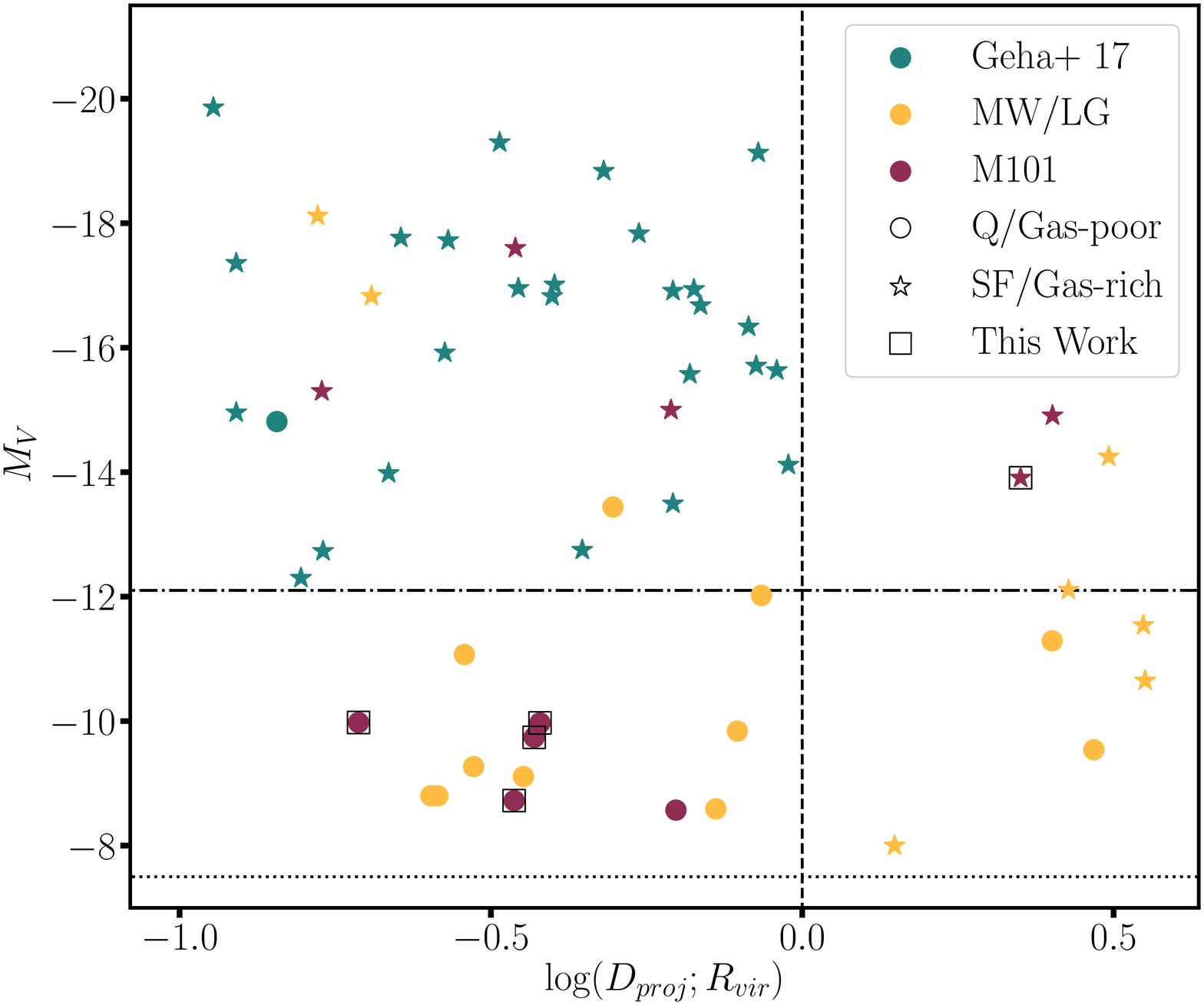}
\caption{$M_V$ as a function of projected separation, $D_{proj}$, at the host distance for all known M101 satellites \citep[maroon symbols;][]{Tikhonov,2017M101LuminosityFn,Paulm101hst} brighter than $M_V=-7.5$, all known Milky Way/Local Group satellites down to that same limit \citep[yellow symbols; 2015 update of][]{lgcatalog}, and satellites brighter than $M_V=-12.1$ from the 8 spectroscopically complete hosts from the SAGA survey \citep[horizontal dash-dotted line,][]{GehaSAGA}. Stars denote satellites that are star forming or gas rich, and circles denote satellites that are quiescent or gas poor. M101 satellites with HI observations from this work are enclosed by black boxes, and the vertical dashed line highlights $D_{proj}=R_{vir}$. We adopt $R_{vir}=260$ kpc for M101 and $R_{vir}=300$ kpc for the Milky Way and SAGA hosts.}
\label{fig:saga}
\end{figure*}

Our observations afford comparisons between the gas content of confirmed M101 satellites with that of the satellites of the Milky Way as well as with the populations around galaxies of similar mass.\ Figure \ref{fig:saga} makes such a comparison: we plot $M_V$ as a function of projected separation, $D_{proj}$, at the host distance for all known M101 satellites \citep[maroon symbols;][]{Tikhonov,2017M101LuminosityFn,Paulm101hst} down to the B17 completeness limit of $M_V = -7.5$ (horizontal dotted line), all known Milky Way/Local Group sattelites down to that same limit \citep[yellow symbols; 2015 update of][]{lgcatalog}, and the 27 satellites detected within the projected virial radii of 8 Milky Way mass hosts from the SAGA survey. The SAGA satellite populations are  spectroscopically complete down to a limit of $M_r = -12.3 \, \mathrm{, or}\, M_V = -12.1$ for a median $g-r=0.4$ \citep[horizontal dash-dotted line,][]{GehaSAGA}. The stars denote satellites that are star forming or gas rich, and the circles denote satellites that are quiescent or gas poor.\ M101 satellites with HI observations from this work are enclosed by black boxes, and the vertical dashed line indicates $D_{proj}=R_{vir}$.\ We adopt $R_{vir}= 260$ kpc for M101 \citep{2014MerritLSBsM101} and $R_{vir}= 300 $ kpc for the Milky Way and the SAGA hosts (see \citealt{GehaSAGA} for more detail).

Fig.\ \ref{fig:saga} illustrates that the satellites within $R_{vir}$ of M101, the Milky Way or the SAGA hosts that are brighter than $M_V \simeq -12$ are typically star forming or gas rich, as are most companions beyond $R_{vir}$. By contrast, the satellites within $R_{vir}$ that are fainter than $M_V \simeq -12$ are typically quiescent or gas poor.\ This difference in gas content and star formation activity between bright and faint satellites likely results from environmental effects: the gas reservoirs of low-mass subhalos are more easily stripped, and their star formation quenched, due to their shallower potential wells relative to high-mass subhalos \citep{Wetzel2015,Emerick2016,RPS-HIdensityProfiles}. Fig.\ \ref{fig:saga} therefore implies that the high fraction of star-forming satellites detected within $R_{vir}$ by SAGA and the low fraction of HI-rich satellites within the $R_{vir}$ of M101 and the Milky Way can be explained by differences in survey sensitivity rather than inherent differences in the star formation activity among satellite populations.\ Instead we find that, when completeness is taken into account, the gas richnesses of the satellite populations of M101, the Milky Way, and the SAGA hosts within $R_{vir}$ are broadly consistent with one another. This suggests environment has a similar effect on the satellite populations of different Milky Way-mass hosts, in line with expectations from simulations \citep[e.g.][]{Environmentalquenching2018}.

\section{Conclusions} \label{sec:Conclusion}
We have presented HI follow-up observations using the GBT along the LOS to 27 optically detected LSB dwarf candidates in the M101 region (see Table \ref{table:maintable}). We find the HI counterparts of 5 targets (Figure \ref{fig:detectspectra}) and derive their gas properties such as HI masses and velocity widths (Table \ref{table:detectiontable}). We find no HI emission along the LOS associated to the remaining 22 (Figure \ref{fig:nondetectspectra}) and place stringent $5\sigma\, \mathrm{single,} \,  25\,\kms$-channel upper limits on their HI masses and gas richnesses (Table \ref{table:upperlimits}).

 Among detections, we find that dw1343+58 is likely a member of the M101 group while Dw26 is far in the background with $D_{HI}\sim 150 $Mpc. We find that the remaining 3 LSB dwarfs with HI detections (DF-5, DwB, and dw1408+56) are likely members of the background NGC~5485 group. We show that detections and non-detections span a similar range in measured colour (Figure \ref{fig:gasrichness-colour}), implying that at these low surface brightnesses, measured colours are not a good indicator of gas content. These findings demonstrate the utility of HI follow-up of optically detected LSB features to constrain their physical properties.

The optical lower distance limits and projected separations of our HI non-detections from the elliptical galaxy NGC~5485 suggest that they all could be satellites of that group, where any gas they once had has been gas stripped by environmental effects (Figure \ref{fig:m101dist}). In this context, the sphere of influence of the NGC~5485 group has a radius $R\sim 1$ Mpc, in-line with observational and theoretical constraints.

We compare the gas richnesses of confirmed M101 satellites to those of the Milky Way satellites and of similar mass hosts in the SAGA survey (Figure \ref{fig:saga}). Accounting for completeness, we find general agreement between these populations: satellites within the virial radius brighter than $M_V \simeq -12$ are broadly star-forming and gas-rich while those fainter than this threshold are broadly quiescent and gas-poor. This suggests that the effect on satellite gas content is similar around hosts of similar stellar mass, in line with theoretical expectations.

\acknowledgments
We thank the anonymous referee for their useful comments on improving the clarity of this paper. KS acknowledges support from the Natural Sciences and Engineering Research Council of Canada (NSERC). Research by DJS is supported by NSF grants AST-1821967, 1821987, 1813708, 1813466, and 1908972. Research by DC is supported by NSF grant AST-1814208, and by NASA through grants number HST-GO-15426.007-A and HST-GO-HST-GO-15332.004-A from the Space Telescope Science Institute, which is operated by AURA, Inc., under NASA contract NAS 5-26555.

The Green Bank Observatory is a facility of the National Science Foundation operated under cooperative agreement by Associated Universities, Inc.

\vspace{5mm}
\facilities{GBT (VEGAS)}

\software{astropy \citep{astropy:2013,astropy:2018}, GBTIDL}

\bibliographystyle{aasjournal}
\bibliography{references}

\begin{thebibliography}{}
\expandafter\ifx\csname natexlab\endcsname\relax\def\natexlab#1{#1}\fi
\providecommand{\url}[1]{\href{#1}{#1}}
\providecommand{\dodoi}[1]{doi:~\href{http://doi.org/#1}{\nolinkurl{#1}}}
\providecommand{\doeprint}[1]{\href{http://ascl.net/#1}{\nolinkurl{http://ascl.net/#1}}}
\providecommand{\doarXiv}[1]{\href{https://arxiv.org/abs/#1}{\nolinkurl{https://arxiv.org/abs/#1}}}

\bibitem[{{Abraham} \& {van Dokkum}(2014)}]{DragonflyArrayOriginal}
{Abraham}, R.~G., \& {van Dokkum}, P.~G. 2014, Publications of the Astronomical
  Society of Pacific, 126, 55, \dodoi{10.1086/674875}

\bibitem[{{Aguado} {et~al.}(2019){Aguado}, {Ahumada}, {Almeida}, {Anderson},
  {Andrews}, {Anguiano}, {Aquino Ort{\'\i}z}, {Arag{\'o}n-Salamanca},
  {Argudo-Fern{\'a}ndez}, {Aubert}, {Avila-Reese}, {Badenes}, {Barboza
  Rembold}, {Barger}, {Barrera-Ballesteros}, {Bates}, {Bautista}, {Beaton},
  {Beers}, {Belfiore}, {Bernardi}, {Bershady}, {Beutler}, {Bird}, {Bizyaev},
  {Blanc}, {Blanton}, {Blomqvist}, {Bolton}, {Boquien}, {Borissova}, {Bovy},
  {Brand t}, {Brinkmann}, {Brownstein}, {Bundy}, {Burgasser}, {Byler}, {Cano
  Diaz}, {Cappellari}, {Carrera}, {Cervantes Sodi}, {Chen}, {Cherinka}, {Choi},
  {Chung}, {Coffey}, {Comerford}, {Comparat}, {Covey}, {da Silva Ilha}, {da
  Costa}, {Dai}, {Damke}, {Darling}, {Davies}, {Dawson}, {de Sainte Agathe},
  {Deconto Machado}, {Del Moro}, {De Lee}, {Diamond-Stanic}, {Dom{\'\i}nguez
  S{\'a}nchez}, {Donor}, {Drory}, {du Mas des Bourboux}, {Duckworth}, {Dwelly},
  {Ebelke}, {Emsellem}, {Escoffier}, {Fern{\'a}ndez-Trincado}, {Feuillet},
  {Fischer}, {Fleming}, {Fraser-McKelvie}, {Freischlad}, {Frinchaboy}, {Fu},
  {Galbany}, {Garcia-Dias}, {Garc{\'\i}a-Hern{\'a}ndez}, {Garma Oehmichen},
  {Geimba Maia}, {Gil-Mar{\'\i}n}, {Grabowski}, {Gu}, {Guo}, {Ha},
  {Harrington}, {Hasselquist}, {Hayes}, {Hearty}, {Hernandez Toledo}, {Hicks},
  {Hogg}, {Holley-Bockelmann}, {Holtzman}, {Hsieh}, {Hunt}, {Hwang},
  {Ibarra-Medel}, {Jimenez Angel}, {Johnson}, {Jones}, {J{\"o}nsson},
  {Kinemuchi}, {Kollmeier}, {Krawczyk}, {Kreckel}, {Kruk}, {Lacerna}, {Lan},
  {Lane}, {Law}, {Lee}, {Li}, {Lian}, {Lin}, {Lin}, {Lintott}, {Long},
  {Longa-Pe{\~n}a}, {Mackereth}, {de la Macorra}, {Majewski}, {Malanushenko},
  {Manchado}, {Maraston}, {Mariappan}, {Marinelli}, {Marques-Chaves},
  {Masseron}, {Masters}, {McDermid}, {Medina Pe{\~n}a}, {Meneses-Goytia},
  {Merloni}, {Merrifield}, {Meszaros}, {Minniti}, {Minsley}, {Muna}, {Myers},
  {Nair}, {Correa do Nascimento}, {Newman}, {Nitschelm}, {Olmstead}, {Oravetz},
  {Oravetz}, {Ortega Minakata}, {Pace}, {Padilla}, {Palicio}, {Pan}, {Pan},
  {Parikh}, {Parker}, {Peirani}, {Penny}, {Percival}, {Perez-Fournon},
  {Peterken}, {Pinsonneault}, {Prakash}, {Raddick}, {Raichoor}, {Riffel},
  {Riffel}, {Rix}, {Robin}, {Roman-Lopes}, {Rose}, {Ross}, {Rossi}, {Rowlands},
  {Rubin}, {S{\'a}nchez}, {S{\'a}nchez-Gallego}, {Sayres}, {Schaefer},
  {Schiavon}, {Schimoia}, {Schlafly}, {Schlegel}, {Schneider}, {Schultheis},
  {Seo}, {Shamsi}, {Shao}, {Shen}, {Shetty}, {Simonian}, {Smethurst}, {Sobeck},
  {Souter}, {Spindler}, {Stark}, {Stassun}, {Steinmetz}, {Storchi-Bergmann},
  {Stringfellow}, {Su{\'a}rez}, {Sun}, {Taghizadeh-Popp}, {Talbot}, {Tayar},
  {Thakar}, {Thomas}, {Tissera}, {Tojeiro}, {Troup}, {Unda-Sanzana},
  {Valenzuela}, {Vargas-Maga{\~n}a}, {V{\'a}zquez-Mata}, {Wake}, {Weaver},
  {Weijmans}, {Westfall}, {Wild}, {Wilson}, {Woods}, {Yan}, {Yang}, {Zamora},
  {Zasowski}, {Zhang}, {Zheng}, {Zheng}, {Zhu}, {Zinn}, \& {Zou}}]{SDSSDR15}
{Aguado}, D.~S., {Ahumada}, R., {Almeida}, A., {et~al.} 2019, \apjs, 240, 23,
  \dodoi{10.3847/1538-4365/aaf651}

\bibitem[{{Aihara} {et~al.}(2018){Aihara}, {Arimoto}, {Armstrong}, {Arnouts},
  {Bahcall}, {Bickerton}, {Bosch}, {Bundy}, {Capak}, {Chan}, {Chiba}, {Coupon},
  {Egami}, {Enoki}, {Finet}, {Fujimori}, {Fujimoto}, {Furusawa}, {Furusawa},
  {Goto}, {Goulding}, {Greco}, {Greene}, {Gunn}, {Hamana}, {Harikane},
  {Hashimoto}, {Hattori}, {Hayashi}, {Hayashi}, {He{\l}miniak}, {Higuchi},
  {Hikage}, {Ho}, {Hsieh}, {Huang}, {Huang}, {Ikeda}, {Imanishi}, {Inoue},
  {Iwasawa}, {Iwata}, {Jaelani}, {Jian}, {Kamata}, {Karoji}, {Kashikawa},
  {Katayama}, {Kawanomoto}, {Kayo}, {Koda}, {Koike}, {Kojima}, {Komiyama},
  {Konno}, {Koshida}, {Koyama}, {Kusakabe}, {Leauthaud}, {Lee}, {Lin}, {Lin},
  {Lupton}, {Mand elbaum}, {Matsuoka}, {Medezinski}, {Mineo}, {Miyama},
  {Miyatake}, {Miyazaki}, {Momose}, {More}, {More}, {Moritani}, {Moriya},
  {Morokuma}, {Mukae}, {Murata}, {Murayama}, {Nagao}, {Nakata}, {Niida},
  {Niikura}, {Nishizawa}, {Obuchi}, {Oguri}, {Oishi}, {Okabe}, {Okamoto},
  {Okura}, {Ono}, {Onodera}, {Onoue}, {Osato}, {Ouchi}, {Price}, {Pyo}, {Sako},
  {Sawicki}, {Shibuya}, {Shimasaku}, {Shimono}, {Shirasaki}, {Silverman},
  {Simet}, {Speagle}, {Spergel}, {Strauss}, {Sugahara}, {Sugiyama}, {Suto},
  {Suyu}, {Suzuki}, {Tait}, {Takada}, {Takata}, {Tamura}, {Tanaka}, {Tanaka},
  {Tanaka}, {Tanaka}, {Terai}, {Terashima}, {Toba}, {Tominaga}, {Toshikawa},
  {Turner}, {Uchida}, {Uchiyama}, {Umetsu}, {Uraguchi}, {Urata}, {Usuda},
  {Utsumi}, {Wang}, {Wang}, {Wong}, {Yabe}, {Yamada}, {Yamanoi}, {Yasuda},
  {Yeh}, {Yonehara}, \& {Yuma}}]{HSC-SurveyOverview}
{Aihara}, H., {Arimoto}, N., {Armstrong}, R., {et~al.} 2018, \pasj, 70, S4,
  \dodoi{10.1093/pasj/psx066}

\bibitem[{{Astropy Collaboration} {et~al.}(2013){Astropy Collaboration},
  {Robitaille}, {Tollerud}, {Greenfield}, {Droettboom}, {Bray}, {Aldcroft},
  {Davis}, {Ginsburg}, {Price-Whelan}, {Kerzendorf}, {Conley}, {Crighton},
  {Barbary}, {Muna}, {Ferguson}, {Grollier}, {Parikh}, {Nair}, {Unther},
  {Deil}, {Woillez}, {Conseil}, {Kramer}, {Turner}, {Singer}, {Fox}, {Weaver},
  {Zabalza}, {Edwards}, {Azalee Bostroem}, {Burke}, {Casey}, {Crawford},
  {Dencheva}, {Ely}, {Jenness}, {Labrie}, {Lim}, {Pierfederici}, {Pontzen},
  {Ptak}, {Refsdal}, {Servillat}, \& {Streicher}}]{astropy:2013}
{Astropy Collaboration}, {Robitaille}, T.~P., {Tollerud}, E.~J., {et~al.} 2013,
  \aap, 558, A33, \dodoi{10.1051/0004-6361/201322068}

\bibitem[{{Bah{\'e}} {et~al.}(2013){Bah{\'e}}, {McCarthy}, {Balogh}, \&
  {Font}}]{2013bahe}
{Bah{\'e}}, Y.~M., {McCarthy}, I.~G., {Balogh}, M.~L., \& {Font}, A.~S. 2013,
  \mnras, 430, 3017, \dodoi{10.1093/mnras/stt109}

\bibitem[{{Bennet} {et~al.}(2019){Bennet}, {Sand}, {Crnojevi{\'c}}, {Spekkens},
  {Karunakaran}, {Zaritsky}, \& {Mutlu-Pakdil}}]{Paulm101hst}
{Bennet}, P., {Sand}, D.~J., {Crnojevi{\'c}}, D., {et~al.} 2019, arXiv
  e-prints, arXiv:1906.03230.
\newblock \doarXiv{1906.03230}

\bibitem[{{Bennet} {et~al.}(2017){Bennet}, {Sand}, {Crnojevi{\'c}}, {Spekkens},
  {Zaritsky}, \& {Karunakaran}}]{Paulspaper}
---. 2017, \apj, 850, 109, \dodoi{10.3847/1538-4357/aa9180}

\bibitem[{{Bradford} {et~al.}(2015){Bradford}, {Geha}, \&
  {Blanton}}]{Bradford2015MAIN}
{Bradford}, J.~D., {Geha}, M.~C., \& {Blanton}, M.~R. 2015, \apj, 809, 146,
  \dodoi{10.1088/0004-637X/809/2/146}

\bibitem[{{Brown} {et~al.}(2015){Brown}, {Catinella}, {Cortese}, {Kilborn},
  {Haynes}, \& {Giovanelli}}]{Brown2015}
{Brown}, T., {Catinella}, B., {Cortese}, L., {et~al.} 2015, \mnras, 452, 2479,
  \dodoi{10.1093/mnras/stv1311}

\bibitem[{{Brown} {et~al.}(2017){Brown}, {Catinella}, {Cortese}, {Lagos},
  {Dav{\'e}}, {Kilborn}, {Haynes}, {Giovanelli}, \&
  {Rafieferantsoa}}]{2017browngasstripping}
---. 2017, \mnras, 466, 1275, \dodoi{10.1093/mnras/stw2991}

\bibitem[{{Carlsten} {et~al.}(2019){Carlsten}, {Beaton}, {Greco}, \&
  {Greene}}]{CarlstenSBFM101}
{Carlsten}, S.~G., {Beaton}, R.~L., {Greco}, J.~P., \& {Greene}, J.~E. 2019,
  \apjl, 878, L16, \dodoi{10.3847/2041-8213/ab24d2}

\bibitem[{{Catinella} {et~al.}(2012){Catinella}, {Schiminovich}, {Kauffmann},
  {Fabello}, {Hummels}, {Lemonias}, {Moran}, {Wu}, {Cooper}, \& {Wang}}]{GASS}
{Catinella}, B., {Schiminovich}, D., {Kauffmann}, G., {et~al.} 2012, \aap, 544,
  A65, \dodoi{10.1051/0004-6361/201219261}

\bibitem[{{Cen} {et~al.}(2014){Cen}, {Roxana Pop}, \&
  {Bahcall}}]{2014cengasloss}
{Cen}, R., {Roxana Pop}, A., \& {Bahcall}, N.~A. 2014, Proceedings of the
  National Academy of Science, 111, 7914, \dodoi{10.1073/pnas.1407300111}

\bibitem[{{Chiboucas} {et~al.}(2013){Chiboucas}, {Jacobs}, {Tully}, \&
  {Karachentsev}}]{M81Confirmed}
{Chiboucas}, K., {Jacobs}, B.~A., {Tully}, R.~B., \& {Karachentsev}, I.~D.
  2013, \aj, 146, 126, \dodoi{10.1088/0004-6256/146/5/126}

\bibitem[{{Chiboucas} {et~al.}(2009){Chiboucas}, {Karachentsev}, \&
  {Tully}}]{M81Groupmembers}
{Chiboucas}, K., {Karachentsev}, I.~D., \& {Tully}, R.~B. 2009, \aj, 137, 3009,
  \dodoi{10.1088/0004-6256/137/2/3009}

\bibitem[{{Crnojevi{\'c}} {et~al.}(2016){Crnojevi{\'c}}, {Sand}, {Spekkens},
  {Caldwell}, {Guhathakurta}, {McLeod}, {Seth}, {Simon}, {Strader}, \&
  {Toloba}}]{PISCeS-CenAsats}
{Crnojevi{\'c}}, D., {Sand}, D.~J., {Spekkens}, K., {et~al.} 2016, \apj, 823,
  19, \dodoi{10.3847/0004-637X/823/1/19}

\bibitem[{Crnojevi{\'{c}} {et~al.}(2019)Crnojevi{\'{c}}, Sand, Bennet, Pasetto,
  Spekkens, Caldwell, Guhathakurta, McLeod, Seth, Simon, Strader, \&
  Toloba}]{Crnojevi__2019}
Crnojevi{\'{c}}, D., Sand, D.~J., Bennet, P., {et~al.} 2019, The Astrophysical
  Journal, 872, 80, \dodoi{10.3847/1538-4357/aafbe7}

\bibitem[{{Danieli} {et~al.}(2017){Danieli}, {van Dokkum}, {Merritt},
  {Abraham}, {Zhang}, {Karachentsev}, \& {Makarova}}]{2017M101LuminosityFn}
{Danieli}, S., {van Dokkum}, P., {Merritt}, A., {et~al.} 2017, \apj, 837, 136,
  \dodoi{10.3847/1538-4357/aa615b}

\bibitem[{{Di Cintio} {et~al.}(2019){Di Cintio}, {Brook}, {Macci{\`o}},
  {Dutton}, \& {Cardona-Barrero}}]{NIHAOLSB}
{Di Cintio}, A., {Brook}, C.~B., {Macci{\`o}}, A.~V., {Dutton}, A.~A., \&
  {Cardona-Barrero}, S. 2019, \mnras, 486, 2535, \dodoi{10.1093/mnras/stz985}

\bibitem[{{El-Badry} {et~al.}(2016){El-Badry}, {Wetzel}, {Geha}, {Hopkins},
  {Kere{\v{s}}}, {Chan}, \& {Faucher-Gigu{\`e}re}}]{FIRE-lowmass}
{El-Badry}, K., {Wetzel}, A., {Geha}, M., {et~al.} 2016, \apj, 820, 131,
  \dodoi{10.3847/0004-637X/820/2/131}

\bibitem[{{Emerick} {et~al.}(2016){Emerick}, {Mac Low}, {Grcevich}, \&
  {Gatto}}]{Emerick2016}
{Emerick}, A., {Mac Low}, M.-M., {Grcevich}, J., \& {Gatto}, A. 2016, \apj,
  826, 148, \dodoi{10.3847/0004-637X/826/2/148}

\bibitem[{{Fielder} {et~al.}(2019){Fielder}, {Mao}, {Newman}, {Zentner}, \&
  {Licquia}}]{2019fieldercosmicvariance}
{Fielder}, C.~E., {Mao}, Y.-Y., {Newman}, J.~A., {Zentner}, A.~R., \&
  {Licquia}, T.~C. 2019, \mnras, 486, 4545, \dodoi{10.1093/mnras/stz1098}

\bibitem[{{Fillingham} {et~al.}(2018){Fillingham}, {Cooper}, {Boylan-Kolchin},
  {Bullock}, {Garrison-Kimmel}, \& {Wheeler}}]{Environmentalquenching2018}
{Fillingham}, S.~P., {Cooper}, M.~C., {Boylan-Kolchin}, M., {et~al.} 2018,
  \mnras, 477, 4491, \dodoi{10.1093/mnras/sty958}

\bibitem[{{Fillingham} {et~al.}(2016){Fillingham}, {Cooper}, {Pace},
  {Boylan-Kolchin}, {Bullock}, {Garrison-Kimmel}, \&
  {Wheeler}}]{RPS-HIdensityProfiles}
{Fillingham}, S.~P., {Cooper}, M.~C., {Pace}, A.~B., {et~al.} 2016, \mnras,
  463, 1916, \dodoi{10.1093/mnras/stw2131}

\bibitem[{{Garrison-Kimmel} {et~al.}(2019){Garrison-Kimmel}, {Wetzel},
  {Hopkins}, {Sanderson}, {El-Badry}, {Graus}, {Chan}, {Feldmann},
  {Boylan-Kolchin}, {Hayward}, {Bullock}, {Fitts}, {Samuel}, {Wheeler},
  {Kere{\v{s}}}, \& {Faucher-Gigu{\`e}re}}]{2019garrisonkimmel}
{Garrison-Kimmel}, S., {Wetzel}, A., {Hopkins}, P.~F., {et~al.} 2019, \mnras,
  489, 4574, \dodoi{10.1093/mnras/stz2507}

\bibitem[{{Gatto} {et~al.}(2013){Gatto}, {Fraternali}, {Read}, {Marinacci},
  {Lux}, \& {Walch}}]{RPS-Gatto2013}
{Gatto}, A., {Fraternali}, F., {Read}, J.~I., {et~al.} 2013, \mnras, 433, 2749,
  \dodoi{10.1093/mnras/stt896}

\bibitem[{{Geha} {et~al.}(2012){Geha}, {Blanton}, {Yan}, \&
  {Tinker}}]{GehaNSAfielddwarfs}
{Geha}, M., {Blanton}, M.~R., {Yan}, R., \& {Tinker}, J.~L. 2012, \apj, 757,
  85, \dodoi{10.1088/0004-637X/757/1/85}

\bibitem[{{Geha} {et~al.}(2017){Geha}, {Wechsler}, {Mao}, {Tollerud}, {Weiner},
  {Bernstein}, {Hoyle}, {Marchi}, {Marshall}, {Mu{\~n}oz}, \& {Lu}}]{GehaSAGA}
{Geha}, M., {Wechsler}, R.~H., {Mao}, Y.-Y., {et~al.} 2017, \apj, 847, 4,
  \dodoi{10.3847/1538-4357/aa8626}

\bibitem[{{Gil de Paz} {et~al.}(2003){Gil de Paz}, {Madore}, \&
  {Pevunova}}]{bcdgil}
{Gil de Paz}, A., {Madore}, B.~F., \& {Pevunova}, O. 2003, \apjs, 147, 29,
  \dodoi{10.1086/374737}

\bibitem[{{Grcevich} \& {Putman}(2009)}]{HILocalGroup}
{Grcevich}, J., \& {Putman}, M.~E. 2009, \apj, 696, 385,
  \dodoi{10.1088/0004-637X/696/1/385}

\bibitem[{{Haynes} \& {Giovanelli}(1984)}]{1984AJ.....89..758H}
{Haynes}, M.~P., \& {Giovanelli}, R. 1984, \aj, 89, 758, \dodoi{10.1086/113573}

\bibitem[{{Huang} {et~al.}(2012){Huang}, {Haynes}, {Giovanelli}, \&
  {Brinchmann}}]{huangalfalafa}
{Huang}, S., {Haynes}, M.~P., {Giovanelli}, R., \& {Brinchmann}, J. 2012, \apj,
  756, 113, \dodoi{10.1088/0004-637X/756/2/113}

\bibitem[{{Javanmardi} {et~al.}(2016){Javanmardi}, {Martinez-Delgado},
  {Kroupa}, {Henkel}, {Crawford}, {Teuwen}, {Gabany}, {Hanson}, {Chonis}, \&
  {Neyer}}]{JavanmardiDGSAT}
{Javanmardi}, B., {Martinez-Delgado}, D., {Kroupa}, P., {et~al.} 2016,
  Astronomy \& Astrophysics, 588, A89, \dodoi{10.1051/0004-6361/201527745}

\bibitem[{{Jester} {et~al.}(2005){Jester}, {Schneider}, {Richards}, {Green},
  {Schmidt}, {Hall}, {Strauss}, {Vand en Berk}, {Stoughton}, \&
  {Gunn}}]{Jestervband}
{Jester}, S., {Schneider}, D.~P., {Richards}, G.~T., {et~al.} 2005, \aj, 130,
  873, \dodoi{10.1086/432466}

\bibitem[{{Karachentsev} {et~al.}(2014){Karachentsev}, {Bautzmann}, {Neyer},
  {Polzl}, {Riepe}, {Zilch}, \& {Mattern}}]{TBG-Kara2014}
{Karachentsev}, I.~D., {Bautzmann}, D., {Neyer}, F., {et~al.} 2014, ArXiv
  e-prints, arXiv:1401.2719.
\newblock \doarXiv{1401.2719}

\bibitem[{{Karachentsev} \& {Makarova}(2019)}]{karaandmakaM101}
{Karachentsev}, I.~D., \& {Makarova}, L.~N. 2019, Astrophysics, 62, 293,
  \dodoi{10.1007/s10511-019-09582-7}

\bibitem[{{Karachentsev} {et~al.}(2015){Karachentsev}, {Riepe}, {Zilch},
  {Blauensteiner}, {Elvov}, {Hochleitner}, {Hubl}, {Kerschhuber},
  {K{\"u}ppers}, {Neyer}, {P{\"o}lzl}, {Remmel}, {Schneider}, {Sparenberg},
  {Trulson}, {Willems}, \& {Ziegler}}]{2015AstBu..70..379K}
{Karachentsev}, I.~D., {Riepe}, P., {Zilch}, T., {et~al.} 2015, Astrophysical
  Bulletin, 70, 379, \dodoi{10.1134/S199034131504001X}

\bibitem[{{Lee} \& {Jang}(2012)}]{LeeandJang2012}
{Lee}, M.~G., \& {Jang}, I.~S. 2012, \apjl, 760, L14,
  \dodoi{10.1088/2041-8205/760/1/L14}

\bibitem[{{McConnachie}(2012)}]{lgcatalog}
{McConnachie}, A.~W. 2012, \aj, 144, 4, \dodoi{10.1088/0004-6256/144/1/4}

\bibitem[{{Merritt} {et~al.}(2014){Merritt}, {van Dokkum}, \&
  {Abraham}}]{2014MerritLSBsM101}
{Merritt}, A., {van Dokkum}, P., \& {Abraham}, R. 2014, \apjl, 787, L37,
  \dodoi{10.1088/2041-8205/787/2/L37}

\bibitem[{{Merritt} {et~al.}(2016){Merritt}, {van Dokkum}, {Danieli},
  {Abraham}, {Zhang}, {Karachentsev}, \& {Makarova}}]{2016ApJ...833..168M}
{Merritt}, A., {van Dokkum}, P., {Danieli}, S., {et~al.} 2016, \apj, 833, 168,
  \dodoi{10.3847/1538-4357/833/2/168}

\bibitem[{{Mihos} {et~al.}(2013){Mihos}, {Harding}, {Spengler}, {Rudick}, \&
  {Feldmeier}}]{m101vsys}
{Mihos}, J.~C., {Harding}, P., {Spengler}, C.~E., {Rudick}, C.~S., \&
  {Feldmeier}, J.~J. 2013, \apj, 762, 82, \dodoi{10.1088/0004-637X/762/2/82}

\bibitem[{{Mihos} {et~al.}(2012){Mihos}, {Keating}, {Holley-Bockelmann},
  {Pisano}, \& {Kassim}}]{Mihosm101HI}
{Mihos}, J.~C., {Keating}, K.~M., {Holley-Bockelmann}, K., {Pisano}, D.~J., \&
  {Kassim}, N.~E. 2012, \apj, 761, 186, \dodoi{10.1088/0004-637X/761/2/186}

\bibitem[{{Moster} {et~al.}(2011){Moster}, {Somerville}, {Newman}, \&
  {Rix}}]{Cosmicvariance}
{Moster}, B.~P., {Somerville}, R.~S., {Newman}, J.~A., \& {Rix}, H.-W. 2011,
  \apj, 731, 113, \dodoi{10.1088/0004-637X/731/2/113}

\bibitem[{{M{\"u}ller} {et~al.}(2018){M{\"u}ller}, {Jerjen}, \&
  {Binggeli}}]{LeoILeoTriplet}
{M{\"u}ller}, O., {Jerjen}, H., \& {Binggeli}, B. 2018, \aap, 615, A105,
  \dodoi{10.1051/0004-6361/201832897}

\bibitem[{{M{\"u}ller} {et~al.}(2017){M{\"u}ller}, {Scalera}, {Binggeli}, \&
  {Jerjen}}]{Mullerspaper}
{M{\"u}ller}, O., {Scalera}, R., {Binggeli}, B., \& {Jerjen}, H. 2017,
  Astronomy \& Astrophysics, 602, A119, \dodoi{10.1051/0004-6361/201730434}

\bibitem[{{Price-Whelan} {et~al.}(2018){Price-Whelan}, {Sip{\H{o}}cz},
  {G{\"u}nther}, {Lim}, {Crawford}, {Conseil}, {Shupe}, {Craig}, {Dencheva},
  {Ginsburg}, {VanderPlas}, {Bradley}, {P{\'e}rez-Su{\'a}rez}, {de Val-Borro},
  {Paper Contributors}, {Aldcroft}, {Cruz}, {Robitaille}, {Tollerud},
  {Coordination Committee}, {Ardelean}, {Babej}, {Bach}, {Bachetti}, {Bakanov},
  {Bamford}, {Barentsen}, {Barmby}, {Baumbach}, {Berry}, {Biscani}, {Boquien},
  {Bostroem}, {Bouma}, {Brammer}, {Bray}, {Breytenbach}, {Buddelmeijer},
  {Burke}, {Calderone}, {Cano Rodr{\'\i}guez}, {Cara}, {Cardoso}, {Cheedella},
  {Copin}, {Corrales}, {Crichton}, {D{\textquoteright}Avella}, {Deil},
  {Depagne}, {Dietrich}, {Donath}, {Droettboom}, {Earl}, {Erben}, {Fabbro},
  {Ferreira}, {Finethy}, {Fox}, {Garrison}, {Gibbons}, {Goldstein}, {Gommers},
  {Greco}, {Greenfield}, {Groener}, {Grollier}, {Hagen}, {Hirst}, {Homeier},
  {Horton}, {Hosseinzadeh}, {Hu}, {Hunkeler}, {Ivezi{\'c}}, {Jain}, {Jenness},
  {Kanarek}, {Kendrew}, {Kern}, {Kerzendorf}, {Khvalko}, {King}, {Kirkby},
  {Kulkarni}, {Kumar}, {Lee}, {Lenz}, {Littlefair}, {Ma}, {Macleod},
  {Mastropietro}, {McCully}, {Montagnac}, {Morris}, {Mueller}, {Mumford},
  {Muna}, {Murphy}, {Nelson}, {Nguyen}, {Ninan}, {N{\"o}the}, {Ogaz}, {Oh},
  {Parejko}, {Parley}, {Pascual}, {Patil}, {Patil}, {Plunkett}, {Prochaska},
  {Rastogi}, {Reddy Janga}, {Sabater}, {Sakurikar}, {Seifert}, {Sherbert},
  {Sherwood-Taylor}, {Shih}, {Sick}, {Silbiger}, {Singanamalla}, {Singer},
  {Sladen}, {Sooley}, {Sornarajah}, {Streicher}, {Teuben}, {Thomas},
  {Tremblay}, {Turner}, {Terr{\'o}n}, {van Kerkwijk}, {de la Vega}, {Watkins},
  {Weaver}, {Whitmore}, {Woillez}, {Zabalza}, \& {Contributors}}]{astropy:2018}
{Price-Whelan}, A.~M., {Sip{\H{o}}cz}, B.~M., {G{\"u}nther}, H.~M., {et~al.}
  2018, \aj, 156, 123, \dodoi{10.3847/1538-3881/aabc4f}

\bibitem[{{Saulder} {et~al.}(2016){Saulder}, {van Kampen}, {Chilingarian},
  {Mieske}, \& {Zeilinger}}]{SaulderNGC5485}
{Saulder}, C., {van Kampen}, E., {Chilingarian}, I.~V., {Mieske}, S., \&
  {Zeilinger}, W.~W. 2016, \aap, 596, A14, \dodoi{10.1051/0004-6361/201526711}

\bibitem[{{Schaefer} {et~al.}(2019){Schaefer}, {Croom}, {Scott}, {Brough},
  {Allen}, {Bekki}, {Bland-Hawthorn}, {Bloom}, {Bryant}, {Cortese}, {Davies},
  {Federrath}, {Fogarty}, {Green}, {Groves}, {Hopkins}, {Konstantopoulos},
  {L{\'o}pez-S{\'a}nchez}, {Lawrence}, {McElroy}, {Medling}, {Owers}, {Pracy},
  {Richards}, {Robotham}, {van de Sande}, {Tonini}, \& {Yi}}]{2019schaefer}
{Schaefer}, A.~L., {Croom}, S.~M., {Scott}, N., {et~al.} 2019, \mnras, 483,
  2851, \dodoi{10.1093/mnras/sty3258}

\bibitem[{{Simpson} {et~al.}(2018){Simpson}, {Grand}, {G{\'o}mez}, {Marinacci},
  {Pakmor}, {Springel}, {Campbell}, \& {Frenk}}]{RPS-SimpsonAuriga}
{Simpson}, C.~M., {Grand}, R. J.~J., {G{\'o}mez}, F.~A., {et~al.} 2018, \mnras,
  478, 548, \dodoi{10.1093/mnras/sty774}

\bibitem[{{Smercina} {et~al.}(2018){Smercina}, {Bell}, {Price}, {D'Souza},
  {Slater}, {Bailin}, {Monachesi}, \& {Nidever}}]{M94sats}
{Smercina}, A., {Bell}, E.~F., {Price}, P.~A., {et~al.} 2018, \apj, 863, 152,
  \dodoi{10.3847/1538-4357/aad2d6}

\bibitem[{{Spekkens} \& {Karunakaran}(2018)}]{HCGUDGs}
{Spekkens}, K., \& {Karunakaran}, A. 2018, \apj, 855, 28,
  \dodoi{10.3847/1538-4357/aa94be}

\bibitem[{{Spekkens} {et~al.}(2013){Spekkens}, {Mason}, {Aguirre}, \&
  {Nhan}}]{GBTbeam}
{Spekkens}, K., {Mason}, B.~S., {Aguirre}, J.~E., \& {Nhan}, B. 2013, \apj,
  773, 61, \dodoi{10.1088/0004-637X/773/1/61}

\bibitem[{{Spekkens} {et~al.}(2014){Spekkens}, {Urbancic}, {Mason}, {Willman},
  \& {Aguirre}}]{HIMilkyWay}
{Spekkens}, K., {Urbancic}, N., {Mason}, B.~S., {Willman}, B., \& {Aguirre},
  J.~E. 2014, \apjl, 795, L5, \dodoi{10.1088/2041-8205/795/1/L5}

\bibitem[{{Springob} {et~al.}(2005){Springob}, {Haynes}, {Giovanelli}, \&
  {Kent}}]{2005ApJS..160..149S}
{Springob}, C.~M., {Haynes}, M.~P., {Giovanelli}, R., \& {Kent}, B.~R. 2005,
  \apjs, 160, 149, \dodoi{10.1086/431550}

\bibitem[{{Thuan} \& {Martin}(1981)}]{BCDsgeneral}
{Thuan}, T.~X., \& {Martin}, G.~E. 1981, \apj, 247, 823, \dodoi{10.1086/159094}

\bibitem[{{Tikhonov} {et~al.}(2015){Tikhonov}, {Lebedev}, \&
  {Galazutdinova}}]{Tikhonov}
{Tikhonov}, N.~A., {Lebedev}, V.~S., \& {Galazutdinova}, O.~A. 2015, Astronomy
  Letters, 41, 239, \dodoi{10.1134/S1063773715060080}

\bibitem[{{Tully}(2015)}]{Galaxygroups2massTully}
{Tully}, R.~B. 2015, \aj, 149, 171, \dodoi{10.1088/0004-6256/149/5/171}

\bibitem[{{van der Burg} {et~al.}(2017){van der Burg}, {Hoekstra}, {Muzzin},
  {Sif{\'o}n}, {Viola}, {Bremer}, {Brough}, {Driver}, {Erben}, {Heymans},
  {Hildebrandt}, {Holwerda}, {Klaes}, {Kuijken}, {McGee}, {Nakajima},
  {Napolitano}, {Norberg}, {Taylor}, \& {Valentijn}}]{VanderburgUDGs}
{van der Burg}, R. F.~J., {Hoekstra}, H., {Muzzin}, A., {et~al.} 2017, \aap,
  607, A79, \dodoi{10.1051/0004-6361/201731335}

\bibitem[{{van Dokkum} {et~al.}(2015){van Dokkum}, {Abraham}, {Merritt},
  {Zhang}, {Geha}, \& {Conroy}}]{UDGsvandokkum}
{van Dokkum}, P.~G., {Abraham}, R., {Merritt}, A., {et~al.} 2015, \apjl, 798,
  L45, \dodoi{10.1088/2041-8205/798/2/L45}

\bibitem[{{Wetzel} {et~al.}(2015){Wetzel}, {Tollerud}, \& {Weisz}}]{Wetzel2015}
{Wetzel}, A.~R., {Tollerud}, E.~J., \& {Weisz}, D.~R. 2015, \apjl, 808, L27,
  \dodoi{10.1088/2041-8205/808/1/L27}

\bibitem[{{Wheeler} {et~al.}(2015){Wheeler}, {O{\~n}orbe}, {Bullock},
  {Boylan-Kolchin}, {Elbert}, {Garrison-Kimmel}, {Hopkins}, \&
  {Kere{\v{s}}}}]{Wheeler2015}
{Wheeler}, C., {O{\~n}orbe}, J., {Bullock}, J.~S., {et~al.} 2015, \mnras, 453,
  1305, \dodoi{10.1093/mnras/stv1691}

\bibitem[{{Zaritsky} {et~al.}(2019){Zaritsky}, {Donnerstein}, {Dey},
  {Kadowaki}, {Zhang}, {Karunakaran}, {Mart{\'\i}nez-Delgado}, {Rahman}, \&
  {Spekkens}}]{SMUDGes}
{Zaritsky}, D., {Donnerstein}, R., {Dey}, A., {et~al.} 2019, The Astrophysical
  Journal Supplement Series, 240, 1, \dodoi{10.3847/1538-4365/aaefe9}

\end{thebibliography}

\end{document}